\documentclass[a4paper,11pt]{article}
\pdfoutput=1
\usepackage[utf8]{inputenc}
\usepackage{jcappub}
\usepackage{natbib}
\setcitestyle{square,numbers}
\usepackage{xcolor}
\usepackage{float}
\usepackage{multirow,array,longtable}
\usepackage{arydshln} 
\usepackage[makeroom]{cancel}
\usepackage{amsfonts}
\usepackage{dcolumn}
\usepackage{bm}
\usepackage{graphicx}
\usepackage{amssymb,amsmath}
\usepackage{multirow}
\usepackage{slashed,xfrac}
\usepackage[makeroom]{cancel}
\usepackage{rotating}
\usepackage{color,url}
\usepackage[colorlinks=true
,urlcolor=blue
,anchorcolor=blue
,citecolor=blue
,filecolor=blue
,linkcolor=blue
,menucolor=blue
,linktocpage=true
,pdfproducer=medialab
,pdfa=true
]{hyperref}
\usepackage{epsfig,psfrag,rotating,soul}
\usepackage{rotfloat}
\usepackage[normalem]{ulem}
\usepackage{framed,fancybox,xcolor,xfrac,geometry}
\usepackage{hhline}
\renewcommand\eqref[1]{Eq.~(\ref{#1})}
\newcommand\eqrefs[2]{Eqs.~(\ref{#1})-(\ref{#2})}
\newcommand\figref[1]{Fig.~\ref{#1}}
\newcommand\figrefs[2]{Figs.~\ref{#1}-\ref{#2}}

\newcommand\secref[1]{Section~\ref{#1}}
\newcommand\appref[1]{Appendix~\ref{#1}}

\evensidemargin \oddsidemargin
\allowdisplaybreaks

\def\gY{g'}
\def\gYc{g^{'\,2}}
\def\cw{c_{\rm w}}
\def\sw{s_{\rm w}}

\def\mw{m_{\rm W}}
\def\mz{m_{\rm Z}}
\def\mh{m_{\rm H}}

\def\vev{{\it v}}

\def\Zf{Z}
\def\div{\Delta_\epsilon}
\def\deltaCT{{\delta_\epsilon}}

\def\gmunu{g^{\mu\nu}}
\def\kVnu{k_1^\nu}
\def\kVmu{k_2^\mu}
\def\kTnu{k_2^\nu}
\def\kTmu{k_1^\mu}

\newcommand{\nn}{\nonumber}
\newcommand{\be}{\begin{equation}}
\newcommand{\ee}{\end{equation}}
\newcommand{\bear}{\begin{eqnarray}}
\newcommand{\eear}{\end{eqnarray}}
\newcommand{\mL}{\mathcal{L}}
\newcommand{\mO}{\mathcal{O}}

\newcommand{\mF}{\mathcal{F}}
\newcommand{\mV}{\mathcal{V}}
\def\1loop{one-loop}

\def\greenfR{\hat{\Gamma}}
\def\SMgreenfR{\hat{\overline{\Gamma}}}
\def\greenfT{\Gamma^{\rm Tree}}
\def\greenfL{\Gamma^{\rm Loop}}
\def\SMgreenfL{\overline{\Gamma}^{\rm Loop}}
\def\greenfC{\Gamma^{\rm CT}}
\def\amp{\mathcal{A}}

\def\propR{\hat{\Delta}}
\def\SER{\hat{\Sigma}}
\def\SMSigma{\overline{\Sigma}}
\def\SERSM{\hat{\overline{\Sigma}}}

\def\treeLtwo{{\rm EChL}^{(2)}_{\rm Tree}}
\def\treeLfour{{\rm EChL}^{(4)}_{\rm Tree}}
\def\treeLtwoLfour{{\rm EChL}^{(2+4)}_{\rm Tree}}
\def\loopLtwo{{\rm EChL}^{(2)}_{\rm Loop}}
\def\full{{\rm EChL}_{\rm Full}}
\def\fullSM{{\rm SM}_{\rm Full}}
\def\treeSM{{\rm SM}_{\rm Tree}}
\def\loopSM{{\rm SM}_{\rm Loop}}

\usepackage{soul}
\usepackage{lineno} 

\title{One-loop corrections for WW to HH in HEFT with the electroweak chiral Lagrangian}
\author[a]{M. J.  Herrero,}
\author[b]{and R. A.  Morales}
\affiliation[a]{Departamento de F\'{\i}sica Te\'orica and Instituto de F\'{\i}sica Te\'orica, IFT-UAM/CSIC,\\
Universidad Aut\'onoma de Madrid, Cantoblanco, 28049 Madrid, Spain}
\affiliation[b]{IFLP, CONICET - Dpto. de F\'{\i}sica, Universidad Nacional de La Plata, \\ 
C.C. 67, 1900 La Plata, Argentina}
\emailAdd{maria.herrero@uam.es}
\emailAdd{roberto.morales@fisica.unlp.edu.ar}

\abstract{In this work we present the computation of the one-loop  electroweak radiative corrections to the scattering process $WW\to HH$ within the context of the Higgs Effective Field Theory (HEFT).  We assume that the fermionic interactions are like in the Standard Model,  whereas the Beyond Standard Model interactions in the bosonic sector are given by the Electroweak Chiral Lagrangian.  The computation of the one-loop amplitude and the renormalization program is performed in terms of the involved  one-particle-irreducible functions (1PI) and using $R_\xi$ covariant gauges.  The renormalization  of 1PI functions at arbitrary external momenta is a more ambitious program than just renormalizing the amplitude with on-shell external legs and it has the advantage that they can be used in several scattering amplitudes.  In fact, we use here some of the 1PI functions already computed in our previous work (devoted to $WZ  \to WZ$).  We will complement them here with the computation of the new 1PI functions required for $WW \to HH$.  From this renormalization procedure  we will also derive the full set of renormalized coefficients of the EChL that are relevant for this scattering process.  In the last part, we will present the numerical results for the EChL predictions of the one-loop level cross section,  $\sigma(WW \to HH)|_{\rm 1-loop}$,  as a function of the center of mass energy,  showing the relative size of the one-loop radiative corrections respect to the tree level prediction in terms of the EChL coefficients.  The  results of the one-loop corrections to $WW \to HH$ for the SM case will be also presented,  for comparison with the EChL case,  following the same computational method, i.e.,  by means of the renormalization of 1PI functions. }

\begin{document}
\begin{flushright}
	IFT-UAM/CSIC-22-80
\end{flushright}
\maketitle

\section{Introduction}
\label{intro}
The use of Effective Field Theories (EFTs) to study the phenomenological implications of anomalous Higgs couplings beyond the Standard Model (SM) of Particle Physics  is nowadays a very common strategy, widely employed, to test at colliders the new Higgs physics implied by those anomalous couplings in a model independent way, namely, without assuming a particular Beyond Standard Model (BSM).  The information of the anomalous Higgs couplings is encoded in a set of effective operators,  built with the SM fields and with the unique requirement of being invariant under the SM gauge symmetry,  $SU(3) \times SU(2) \times U(1)$.  The coefficients in front of these operators (usually called Wilson coefficients) are generically  unknown and  encode the information of the particular underlying fundamental theory that generates such EFT at low energies when the new heavy modes of this theory are integrated out.  Depending on the kind of dynamics involved in the fundamental theory,  it is more appropriate the use of one EFT or another.  Usually, the so-called SMEFT (Standard Model Effective Theory) is more appropriate to describe the low energy behaviour of weakly interacting dynamics,  whereas the so-called HEFT (Higgs Effective Field Theory) is more appropriate to describe strongly interacting underlying dynamics (for  reviews,  see for instance,  \cite{Brivio:2017vri, Dobado:2019fxe}).  Here we choose this second case,  the HEFT,  and focus on the bosonic sector which is described in terms of the so-called Electroweak Chiral Lagrangian (EChL).  The fermionic sector will be assumed to be as in the SM,  so that no BSM interactions nor effective operators are considered in the fermionic sector of the HEFT. 

Our main goal here is to determine,  within this EChL context,  the size of the one-loop electroweak (EW) radiative corrections for the subprocess at colliders,  $W^+W^- \to HH$, where  two Higgs bosons are produced from the scattering of  two $W$ gauge bosons which are radiated from the initial colliding particles (also called $WW$ fusion in the literature).
From now on,  for brevity,  we omit the explicit charges of the $W$ bosons.
This $WW$ scattering subprocess is known to be  relevant for both types of colliders,  $e^+e^-$ and $pp$,  with energies at the TeV domain.  We also wish to compare in this work  these EW radiative corrections in the EChL context with the corresponding ones of $WW\to HH$ within the SM context, thus we do here the two computations in parallel.  Our calculations of the amplitudes and corresponding cross sections in both cases,  the EChL and the SM,  are full bosonic one-loop computations,   including all kind of diagrams in the loops,  and are valid for physical $W$ and $H$ particles in the external legs,  with all possible polarizations for the $W$ gauge bosons, longitudinal and transverse.  That means that we do not make any approximation for the external legs,  and do not use the Equivalence Theorem which  replaces the external $W_L$'s by the corresponding Goldstone bosons (GBs) and is valid only at high energies, $\sqrt{s}\gg\mw$.  Our computation of the radiative corrections within the HEFT is therefore valid at all energies,  from the low energies just above the two Higgs boson threshold production,  $2\mh\sim 250$ GeV,  up to the typical EFT scale which,  in the EChL framework,  is set by $4 \pi \vev\sim 3$ TeV with $v=246$ GeV.  

Regarding the technicalities involved in the present computation we follow closely our previous work in Ref. \cite{Herrero:2021iqt} which was addressed to the case of $WZ \to WZ$ scattering.  Concretely,  we follow the standard  Feynman diagrammatic approach and describe the full renormalization program also in terms of \1loop Feynman diagrams.  We organize this computation in terms of the involved one-particle-irreducible (1PI) Green functions, with two, three and four external legs,  and  perform  the renormalization program of these Green functions in covariant  gauges.  As we showed in our previous work \cite{Herrero:2021iqt},  the renormalization of the EChL coefficients must be gauge invariant and therefore independent of the $\xi$ parameter of the $R_\xi$ covariant gauges.  This is an important point of doing the analytical computation of the amplitude in covariant gauges.  Regarding the numerical evaluations for the $WW \to HH$ scattering we will choose here  in particular the Feynman 't Hooft gauge with $\xi=1$.  The renormalization conditions are also fixed here as in Ref. \cite{Herrero:2021iqt},  using  the on-shell scheme for the EW parameters,  like boson masses, $\mw$, $\mz$, $\mh$,  and gauge couplings, and the $\overline{MS}$ scheme for the EChL coefficients.   Our work presented here of the full one-loop corrections for $WW \to HH$ in the EChL is the most complete one in the literature, and improves the previous related works in the literature in various aspects.  The first computation of $WW \to HH$ in~\cite{ Delgado:2013hxa}  was done just for the case with external longitudinal $W$ bosons,  replaced  by external GBs by using the ET,  and include only scalar particles both in the external legs and in the loops,  working always with massless GBs.  A more recent computation of the one-loop radiative corrections for  $WW \to HH$ in the EChL context, in~\cite{Asiain:2021lch},  also refers to the case of longitudinal $W$'s and also uses the ET that replace the external $W_L$ by the GBs which are taken massless.   They consider all kinds of loops for the GBs scattering and compute them in the Landau gauge.  They make the additional approximation of taking equal the $W$ and $Z$ boson masses (called isospin limit in the literature).  Our best improvement respect to these works is that we do not use the ET,  i.e. we work with external gauge bosons, instead of GBs,  we do not take equal masses for W and Z, and we do not take massless GBs,  since we work in the Feynman 't Hooft gauge.  Consequently,  the set of Feynman 1-loop diagrams considered here and in \cite{Delgado:2013hxa, Asiain:2021lch} are also different.  Another important aspect, that we cover in a different way  than in those references is the renormalization program, that we implement here in terms of general renormalized Green functions,  with generic external momenta,  in contrast to Refs. \cite{Delgado:2013hxa, Asiain:2021lch} that apply the renormalization program directly to the on-shell scattering amplitude.  The advantage of doing renormalization at the more general off-shell Green functions level, is that these same renormalized functions can be used as well  for the computation of radiative corrections in another observables.  For instance, we have used the same renormalized vertex function $WWH$ here for $WW \to HH$ than  in our previous computation in Ref. \cite{Herrero:2021iqt} for $WZ \to WZ$.  The difference is just in the particular setting of the external legs momenta of the vertex function which must be done properly for each case.  On the other hand,  the renormalization program using 1PI Green functions instead of just on-shell amplitudes requires the renormalization of a larger set of EChL coefficients.  It is, therefore,  also more complete in this sense.  Due to the relevance of this latter issue,  we will devote some part of the present work to the comparison of our results on the renormalization of the EChL coefficients with some related previous results \cite{Delgado:2013hxa, Asiain:2021lch,Gavela:2014uta,Buchalla:2020kdh}. 

The paper is organized as follows.  
In \secref{sec-EChL}, we briefly describe the main features of the EChL with $R_\xi$ gauge-fixing and set the relevant operators for  the $WW\to HH$ scattering process.  
The diagrammatic computation by means of the 1PI functions is presented in \secref{diag-1pi}.  
The \secref{sec-renorm} is devoted to the renormalization program, including the prescriptions for regularization and renormalization assumed and the summary of all the divergent counterterms.   
The numerical predictions for this observable within the EChL and the SM are presented and discussed  in \secref{sec-plots}.  Finally, we conclude in \secref{sec-conclu}. 


\section{Relevant part of the electroweak chiral Lagrangian}
\label{sec-EChL}
In this section we introduce the part of the bosonic EChL that is needed for the present computation of the EW radiative corrections of the $WW \to HH$ scattering, and  provide some necessary notation.  
In the EChL context,  the active fields  are the EW gauge bosons, $B_\mu$ and $W^a_\mu$ ($a=1,2,3$), their corresponding GBs $\pi^a$ ($a=1,2,3$), and the Higgs boson $H$. The unique requirement for the building of the EChL is the invariance under the EW gauge,  $SU(2)_L\times U(1)_Y$,  transformations.  On the other hand,  the scalar sector of the EChL has an additional invariance under  the EW chiral  $SU(2)_L \times SU(2)_R$,  transformation.
Under this EW chiral transformation the GBs transform non-linearly.  This peculiarity implies multiple GBs interactions among themselves and also with the other fields. The Higgs boson field, in contrast,  is invariant under all transformations. 
Usually the GBs are introduced in a non-linear representation via the exponential parametrization, by means of the matrix $U$, which transforms linearly under the EW chiral transformations:
\be 
U(\pi^a) = e^{i \pi^a \tau^a/\vev} \, \, , 
\label{expo}
\ee
where, $\tau^a$, $a=1,2,3$,  are the Pauli matrices and $v=246$ GeV.  
On the other hand, the Higgs field is a singlet of the EW chiral symmetry and the EW gauge symmetry. Hence the interactions of $H$  are introduced via generic polynomials since there are not limitations from symmetry arguments on the implementation of this field and its interactions with itself and with the other fields, in contrast to linear EFTs as the SMEFT.
Finally, the EW gauge bosons are  introduced by the gauge invariance principle.  Thus, they appear in the following pieces of the EChL:
\bear
\hat{B}_\mu &=& \gY B_\mu \tau^3/2\,, \quad \hat{B}_{\mu\nu} = \partial_\mu \hat{B}_\nu -\partial_\nu \hat{B}_\mu \,,  \nn\\
\hat{W}_\mu &=& g W^a_\mu \tau^a/2\,, \quad \hat{W}_{\mu\nu} = \partial_\mu \hat{W}_\nu - \partial_\nu \hat{W}_\mu + i  [\hat{W}_\mu,\hat{W}_\nu ] \,,  \nn\\
D_\mu U &=& \partial_\mu U + i\hat{W}_\mu U - i U\hat{B}_\mu \,, \quad \mV_\mu=(D_\mu U)U^\dagger \,,  \quad {\cal D}_\mu O=\partial_\mu O+i[\hat{W}_\mu,O]\, . 
\eear

As it is usual, the chiral counting arrange the effective operators in the EChL into terms with increasing chiral dimension.  The most relevant ones are  the leading order Lagrangian,  with chiral dimension two,  $\mL_2$, and the next to leading order one with chiral dimension four,  $\mL_4$.  The relevant EChL for the present computation can then be summarized as follows:
\be
{\cal L}_{\rm EChL}=\mL_2+ \mL_4 \,,
\label{EChL}
\ee
In this chiral dimension counting,  it is important to keep in mind that all derivatives and masses count as momentum,  namely, 
$\partial_\mu \,,\,\mw \,,\,\mz \,,\,\mh \,,\,g\vev \,,\,\gY\vev \,,\lambda v\, \sim \mO(p)$. 

Firstly,  the leading order Lagrangian, $\mL_2$ is given by,
\bear
\mL_2 &=& \frac{v^2}{4}\left(1+2a\frac{H}{v}+b\left(\frac{H}{v}\right)^2 +\ldots\right){\rm Tr}\Big[ 
 D_\mu U^\dagger D^\mu U \Big]+\frac{1}{2}\partial_\mu H\partial^\mu H-V(H)  \nn\\
&&-\frac{1}{2g^2} {\rm Tr}\Big[ \hat{W}_{\mu\nu}\hat{W}^{\mu\nu}\Big]
-\frac{1}{2\gYc}{\rm Tr}\Big[ 
\hat{B}_{\mu\nu}\hat{B}^{\mu\nu}\Big]  +\mL_{GF} +\mL_{FP} \,.
\label{eq-L2}
\eear
Here $V(H)$ is the Higgs potential, $\mL_{GF}$ and $\mL_{FP}$,  are the gauge-fixing  Lagrangian and Fadeev-Popov  Lagrangian,  respectively.
From now on,  the dots in the presentation of the relevant pieces in the EChL stand for terms that do not enter in our processes of interest,  $WW \to HH$,  neither at tree level nor at \1loop level and then we omit them. 
The Higgs potential in $\mL_2$   is given by:
\be
V(H) = (-\mu^2+\lambda\vev^2)\vev H +\frac{1}{2}(-\mu^2+3\lambda\vev^2) H^2 +\kappa_3\lambda\vev H^3+\kappa_4\frac{\lambda}{4}H^4 \,.
\ee
For the posterior discussion on renormalization in this EChL context,  it is convenient to define $\mh^2= -\mu^2+3\lambda\vev^2$, then we can eliminate the $\mu^2$ parameter in terms of $\mh^2$. In this case, the linear term (Higgs tadpole) can be simply written as,
\be
T=(\mh^2-2\lambda\vev^2)\vev \,,
\label{tadpoledefinition}
\ee
and the minimum of the potential, corresponding to vanishing tadpole, sets $\mh^2=2\lambda\vev^2$.

Here,  we quantize the EChL as in our previous work \cite{Herrero:2021iqt}, i.e.,  using the linear covariant $R_\xi$ gauges~\cite{Fujikawa:1972fe} with the gauge-fixing Lagrangian given by, 
\be
\mL_{\rm GF} = -F_+F_- -\frac{1}{2}F_{Z}^2 -\frac{1}{2}F_{ A}^2 \,,
\label{GF-lag}
\ee
and  the gauge-fixing functions given by:
\bear
F_\pm=\frac{1}{\sqrt{\xi}}(\partial^{\mu}W_{\mu}^\pm-\xi \mw \pi^\pm) \,,\quad F_Z=\frac{1}{\sqrt{\xi}}(\partial^{\mu}Z_{\mu}-\xi \mz \pi^{3}) \,,\quad F_A=\frac{1}{\sqrt{\xi}}(\partial^{\mu}A_{\mu}) \,.
\label{gaugefixingfunctions}
\eear
Here $\xi$ is the generic gauge-fixing parameter of the $R_\xi$ gauges.  Some comments about the $\xi$ dependence are worth to be added here.  Notice that in our renormalization program,  we demand the renormalization of all the 1PI functions involved at arbitrary momentum for the external legs,  and not just the finitess of the one-loop scattering amplitude.  Thus,  in order to demonstrate explicitly the gauge invariance of the renormalized EChL coefficients, i.e.,  to check that these are $\xi$ independent,   the computation of the loop diagrams involved in the 1PI functions should be done for arbitrary $\xi$ parameter, as it was done in our previous work \cite{Herrero:2021iqt} devoted to $WZ \to WZ$ scattering.   All the final scattering amplitudes with external on-shell particles are of course finite and gauge invariant,  but the involved 1PI functions are $\xi$ dependent,  so the cancellation of the $\xi$ dependence in the final one-loop amplitude is an excellent check of the computation.  In the present paper, for all the numerical estimates of $WW \to HH$ we will choose  in particular the Feynman 't Hooft gauge and, accordingly,  for definiteness, we set $\xi=1$ in the presentation of all our results.

From this previous $R_\xi$ gauge-fixing Lagrangian,  one derives as usual,  the corresponding Faddeev-Popov Lagrangian~\cite{Faddeev:1967fc},  given by:
\be
\mL_{\rm FP} = \sum_{i,j=+,-,Z,A} \bar{c}^{i} \frac{\delta F_i}{\delta \alpha_j} c^j \,,
\label{FP-lag}
\ee
where $c^j$ are the ghost fields and $\alpha_j$ ($j=+,-, Z,A$) are the corresponding gauge transformation parameters
under the local transformations $SU(2)_L\times U(1)_Y$ given by $L=e^{ig\vec{\tau}\cdot\vec{\alpha}_L(x)/2}$ and $R=e^{i\gY\tau^3\alpha_Y(x)/2}$.

Formally, the expressions in \eqrefs{GF-lag}{FP-lag} and the gauge bosons field transformations are the same as in the SM.
However, the scalar transformations in this non-linear EFT differ from the corresponding ones in the SM.
This particularity yields to the absence of interactions among the Higgs boson and ghost fields,  and the presence of new interactions with multiple GBs and two ghost fields. 

Secondly,  the relevant chiral dimension four Lagrangian  for the computation of all the one-loop 1PI functions involved in the $WW \to HH$ scattering amplitude is given by: 
\bear
{\mL}_{4}&=& -a_{dd\mV\mV 1} \frac{\partial^\mu H\,\partial^\nu H}{v^2} {\rm Tr}\Big[ {\cal V}_\mu {\cal V}_\nu \Big] 
-a_{dd\mV\mV 2} \frac{\partial^\mu H\,\partial_\mu H}{v^2} {\rm Tr}\Big[ {\cal V}^\nu {\cal V}_\nu \Big]  \nn\\
&&+\left(a_{11}+a_{H11}\frac{H}{\vev}+a_{HH11}\frac{H^2}{\vev^2}\right) {\rm Tr}\Big[{\cal D}_\mu {\cal V}^\mu {\cal D}_\nu {\cal V}^\nu\Big]  \nn\\
&& -
\frac{\mh^2}{4}\left(2a_{H\mV\mV}\frac{H}{v}+a_{HH\mV\mV}\frac{H^2}{\vev^2} \right){\rm Tr}\Big[ {\cal V}^\mu {\cal V}_\mu \Big]  \nn\\
&& - \left(a_{HWW} \frac{H}{v} +a_{HHWW} \frac{H^2}{v^2}\right) {\rm Tr}\Big[\hat{W}_{\mu\nu} \hat{W}^{\mu\nu}\Big] +i\left(a_{d2} +a_{Hd2}\frac{H}{v}\right)\frac{\partial^\nu H}{v} {\rm Tr}\Big[ \hat{W}_{\mu\nu} {\cal V}^\mu\Big]  \nn\\
&&  +\left(a_{\Box\mV\mV} +a_{H\Box\mV\mV}\frac{H}{\vev}\right)\frac{\Box H}{\vev} {\rm Tr}\Big[{\cal V}_\mu {\cal V}^\mu\Big] +\left(a_{d3}+a_{Hd3}\frac{H}{\vev}\right)\frac{\partial ^\nu H}{\vev} {\rm Tr}\Big[{\cal V}_\nu {\cal D}_\mu {\cal V}^\mu \Big]   \nn\\
&&+\left(a_{\Box\Box} +a_{H\Box\Box}\frac{H}{\vev}\right)\frac{\Box H\,\Box H}{\vev^2} +a_{dd\Box}\frac{\partial^\mu H\,\partial_\mu H\,\Box H}{\vev^3}  \nn\\
&&+ \left(a_{Hdd}\frac{\mh^2}{\vev^2}+a_{ddW}\frac{\mw^2}{\vev^2}+a_{ddZ}\frac{\mz^2}{\vev^2}\right)\frac{H}{\vev}\partial^\mu H\,\partial_\mu H 
\label{eq-L4-without-eoms}
\eear

These relevant effective operators are taken from the full HEFT Lagrangian in Ref. \cite{Brivio:2013pma,Gavela:2014uta}, 
 but we use here a different notation for  the EChL coefficients in $\mL_4$,  that are referred here generically as $a_i$'s.   The correspondence among the two set of coefficients,  $a_i$' here and the coefficients in Ref. \cite{Brivio:2013pma,Gavela:2014uta} can be summarized,  in short,  by: 
 $a_{dd\mV\mV 1}\leftrightarrow c_8$, $a_{dd\mV\mV 2}\leftrightarrow c_{20}$, $a_{11}\leftrightarrow c_9$, $a_{HWW}\leftrightarrow a_W$, $a_{HHWW}\leftrightarrow b_W$, $a_{d2}\leftrightarrow c_5$, $a_{Hd2}\leftrightarrow a_5$, $a_{\Box\mV\mV}\leftrightarrow c_7$, $a_{H\Box\mV\mV}\leftrightarrow a_7$, $a_{d3}\leftrightarrow c_{10}$, $a_{Hd3}\leftrightarrow a_{10}$, $a_{\Box\Box}\leftrightarrow c_{\Box H}$, $a_{H\Box\Box}\leftrightarrow a_{\Box H}$, $a_{dd\Box}\leftrightarrow c_{\Delta H}$, $a_{H\mV\mV}\leftrightarrow a_{C}$ and $a_{HH\mV\mV}\leftrightarrow b_{C}$.
 
It is worth noticing that, in contrast to our one-loop computation here,  for a tree level computation of the scattering amplitude (see for instance \cite{ RoberMariaDaniMJ}) the previous set of operators in $\mL_4$ can be reduced to a smaller set by the use of the equations of motion (EOMs).  Concretely, 
one may use the following EOMs involving the pieces $\Box H$ and ${\cal D}_\mu \mV^\mu$:
\bear
\Box H &=& -\frac{\delta V(H)}{\delta H} -\frac{\vev^2}{4}\frac{\mF(H)}{\delta H}{\rm Tr}\Big[ {\cal V}^\mu {\cal V}_\mu \Big] \Rightarrow  \nn\\
&\Rightarrow& \Box H = -\mh^2 H -\frac{3}{2}\kappa_3\mh^2\frac{H^2}{\vev} -\frac{a}{2}\vev{\rm Tr}\Big[ {\cal V}^\mu {\cal V}_\mu \Big] -\frac{b}{2}H{\rm Tr}\Big[ {\cal V}^\mu {\cal V}_\mu \Big]  \nn\\
{\rm Tr}\Big[ \tau^j{\cal D}_\mu \mV^\mu \Big] \mF(H) &=& -{\rm Tr}\Big[ \tau^j \mV^\mu \Big] \partial_\mu\mF(H) \Rightarrow  \nn\\
&\Rightarrow& {\rm Tr}\Big[ \tau^j{\cal D}_\mu \mV^\mu \Big] = -{\rm Tr}\Big[ \tau^j \mV^\mu \Big] \frac{2a}{\vev}\partial_\mu H .
\label{ourgeneral-eoms}
\eear
Then,  one may eliminate  the terms in $\mL_4$ involving these two pieces,  by re-writing them in terms of the other effective operators. This reduces in practice the number of effective operators in $\mL_4$ as follows:
\bear
{\mL}_{4}^{\rm +EOMs}&=& -(a_{dd\mV\mV 1}-4a^2a_{11}+2aa_{d3}) \frac{\partial^\mu H\,\partial^\nu H}{v^2} {\rm Tr}\Big[ {\cal V}_\mu {\cal V}_\nu \Big] -(a_{dd\mV\mV 2}+\frac{a}{2}a_{dd\Box}) \frac{\partial^\mu H\,\partial_\mu H}{v^2} {\rm Tr}\Big[ {\cal V}^\nu {\cal V}_\nu \Big]  \nn\\
&& -\frac{\mh^2}{2}\left(a_{H\mV\mV}-2a_{\Box\mV\mV}+2aa_{\Box\Box}\right)\frac{H}{v} {\rm Tr}\Big[ {\cal V}^\mu {\cal V}_\mu \Big]  \nn\\
&& -\frac{\mh^2}{4}\left(a_{HH\mV\mV}-6\kappa_3a_{\Box\mV\mV}-4a_{H\Box\mV\mV}+4ba_{\Box\Box}+6\kappa_3aa_{\Box\Box}+4aa_{H\Box\Box}\right)\frac{H^2}{\vev^2} {\rm Tr}\Big[ {\cal V}^\mu {\cal V}_\mu \Big]  \nn\\
&&+(a_{Hdd}-a_{dd\Box})\frac{\mh^2}{\vev^2}\frac{H}{\vev}\partial^\mu H\,\partial_\mu H + \left(a_{ddW}\frac{\mw^2}{\vev^2}+a_{ddZ}\frac{\mz^2}{\vev^2}\right)\frac{H}{\vev}\partial^\mu H\,\partial_\mu H  \nn\\
&& - \left(a_{HWW} \frac{H}{\vev} +a_{HHWW} \frac{H^2}{v^2}\right) {\rm Tr}\Big[\hat{W}_{\mu\nu} \hat{W}^{\mu\nu}\Big] +i\left(a_{d2} +a_{Hd2}\frac{H}{v}\right)\frac{\partial^\nu H}{v} {\rm Tr}\Big[ \hat{W}_{\mu\nu} {\cal V}^\mu\Big]
\label{eq-L4-con-eoms-completa}
\eear
In this reduced Lagrangian,  $a_{dd\mV\mV 1}$,  $a_{11}$ and $a_{d3}$ have been grouped together in the first operator,  $a_{dd\mV\mV 2}$ and $a_{dd\Box}$ in the second operator, etc.   We then redefine these combinations of coefficients as follows:
\bear
\eta = \tilde{a}_{dd\mV\mV 1} &\equiv & a_{dd\mV\mV 1}-4a^2a_{11}+2aa_{d3}   \nn\\
\delta = \tilde{a}_{dd\mV\mV 2} &\equiv & a_{dd\mV\mV 2}+\frac{a}{2}a_{dd\Box}  \nn\\
\tilde{a}_{H\mV\mV} &\equiv & a_{H\mV\mV}-2a_{\Box\mV\mV}+2aa_{\Box\Box}  \nn\\
\tilde{a}_{HH\mV\mV} &\equiv & a_{HH\mV\mV}-6\kappa_3a_{\Box\mV\mV}-4a_{H\Box\mV\mV}+4ba_{\Box\Box}+6\kappa_3aa_{\Box\Box}+4aa_{H\Box\Box}  \nn\\
\tilde{a}_{Hdd} &\equiv & a_{Hdd}-a_{dd\Box}  
\label{delta-eta}
\eear
Notice, that in the two first coefficients we have also used the alternative notation given by $\eta$ and $\delta$, which is the one frequently used in some related literature. 
In particular,  the contributions  to the scattering amplitude $WW \to HH$ of these two first operators in the above list  with $\eta$ and $\delta$ coefficients,   were studied firstly in \cite{Delgado:2013hxa,Asiain:2021lch}.  We will compare our results with these and more references in the next sections. 
Notice also that some coefficients, like $a_{H11}$, $a_{HH11}$, $a_{Hd3}$,  have disappeared in \eqref{eq-L4-con-eoms-completa} since,   after the use of the EOMs,  they contribute to effective operators with at least three Higgs bosons which do not enter in the process of our interest here. 

The main consequence of using the EOMs when computing the scattering amplitude is,  that these combinations of the EChL coefficients are the ones appearing precisely in the on-shell scattering amplitudes.  On the other hand,  this means that only these combinations of coefficients are the ones that are really testable at colliders via this particular $WW \to HH$ scattering.  In particular,   only $\eta$ and $\delta$ in Eq. \ref{delta-eta}  and not the separate coefficients,  $a_{dd\mV\mV 1}$,  $a_{11}$, $a_{d3}$,   $a_{dd\mV\mV 2}$, and $a_{dd\Box}$  are the appropriate parameters for a phenomenological analysis of this scattering $WW \to HH$ process.  Similarly for the other combinations appearing in  \eqref{eq-L4-con-eoms-completa}.   However, for our most ambitious computation and renormalization program,  where the finite renormalized one-loop scattering amplitude is obtained in terms of finite renormalized one-loop 1PI functions,  this reduced Lagrangian is not sufficient and we must use the full Lagrangian in Eq. \ref{eq-L4-without-eoms}.  As we will see in the following sections, this full Lagrangian provides not only a finite one-loop amplitude with on-shell external particles but also finite one-loop 1PI functions with arbitrary external momenta (generically off-shell).  This renormalization program in terms of one-loop 1PI functions is also relevant for the check of the gauge invariance of the final one-loop amplitude, and to demonstrate the gauge invariance of the renormalized EChL coefficients.  The great advantage of using this procedure by means of  the 1PI functions to compute the radiative corrections in scattering amplitudes is that the same 1PI one-loop renormalized functions,  once computed at arbitrary external momenta,  can be used for several processes, by just adjusting the external momenta to the ones of that particular process, including the proper on-shell setting for the external legs when needed.  For instance,  the one-loop 1PI function  $\greenfR_{HWW}$ can be used for both $WW \to HH$ and $WZ \to WZ$,  the  one-loop 1PI function $\greenfR_{HHH}$ can be used for both $WW \to HH$ and $HH \to HH$,  and similarly with other processes.  Therefore, our renormalization program based on 1PI functions is more powerful than just to renormalize concrete scattering amplitudes. 

Finally, to end this section we remind that in order to reach the SM tree level vertices from the above presented EChL, one has to set the EFT coefficients to the following reference values: 1) the coefficients in $\mL_2$,  $a$, $b$, $\kappa_3$ and $\kappa_4$ should be set to 1,  and 2) all the coefficients $a_i$ 's in $\mL_4$ should be set to zero.    Accordingly,  the new BSM physics encoded in the EChL is parametrized in terms of the departures from these reference parameter values.  The corresponding derived Feynman Rules (FRs)  from this EChL that are needed for the present computation, together with the corresponding FRs within the SM, were provided in  our previous work,  concretely in Appendix A of~\cite{Herrero:2021iqt},  so we do not repeat them here.  

\section{Diagrammatic computation using  1PI Green functions}
\label{diag-1pi}
In this section we present our procedure for the  computation of the radiative corrections to the amplitude $\amp(WW\to HH)$ by means of the 1PI Green functions.  We apply this procedure to both cases, the EChL and the SM.  

Within the EChL formalism,  the full \1loop scattering amplitude can be splitted into two parts corresponding to the leading order (LO), ${\cal O}(p^2)$, and the next-to-leading order contributions (NLO), ${\cal O}(p^4)$,  which are denoted as $\amp^{(0)}$ and $\amp^{(1)}$ respectively, yielding to
\be
\amp^{\full}\equiv \amp(WW\to HH)^{{\rm EChL}} = \amp^{(0)}(WW\to HH)  + \amp^{(1)}(WW\to HH) \,.
\label{EChLfull}
\ee
In this EChL context, the LO amplitude comes from ${\cal L}_2$ at the tree level, and the NLO one receives typically two contributions.  One contribution comes from  ${\cal L}_4$ at the tree level and the other one comes from the loops computed with ${\cal L}_2$. Thus,  these LO and NLO contributions are written generically as,
\bear
 \amp^{(0)}(WW\to HH)&\equiv &  \amp^{\treeLtwo}\,,    \\
 \amp^{(1)}(WW\to HH)& \equiv &  \amp^{\treeLfour}+ \amp^{\loopLtwo}\,.
\label{LOandNLOamplitudes}
\eear
 The one-loop amplitude in the EChL can also be written in an alternative way,  accounting for the quantum corrections expansion, i.e.,  in powers of $\hbar$. Then,  the full one-loop amplitude is written as:
\be
 \amp^{\full}= \amp^{\treeLtwoLfour}+\amp^{\loopLtwo} \,, 
\label{tree-oneloop}
\ee
where, the tree level amplitude, ${\cal O} (\hbar^0)$,  has contributions from both $\mL_2$ and $\mL_4$,   generically written as $\amp^{\treeLtwoLfour}=\amp^{\treeLtwo}+ \amp^{\treeLfour}$, whereas  the \1loop correction, ${\cal O} (\hbar^1)$,  is obtained  by computing  loops with just $\mL_2$,   generically written as, $\amp^{\loopLtwo}$.  Remember that within the EChL framework,  the $a_i$ coefficients in $\mL_4$,  have a double role and will act as well as counterterms of the extra divergences generated by these loops which can not be absorbed by just the renormalization of the parameters in $\mL_2$. 

On the other hand, we wish to compare the EChL predictions with the SM ones using the same procedure of 1PI functions.  Thus, we will present in parallel the two predictions for the EChL and SM cases.  To our knowledge, our SM computation is the first full bosonic \1loop computation of $WW\to HH$ scattering  using the $R_\xi$ gauges in the literature.
This SM amplitude is defined as the sum of the LO contribution,  which in this case,  is the tree level contribution of ${\cal O} (\hbar^0)$, and the NLO contribution,  which is of ${\cal O} (\hbar^1)$:
\be
\amp^{\fullSM}=\amp^{\treeSM}+ \amp^{\loopSM}\,. 
\label{SMfull}
\ee

For the technical description  of the one-loop radiative corrections  we then organize the \1loop full amplitude  in terms of the 1PI Green functions as follows:
\be
\amp^{\full}=\amp^{\rm LO}+\amp_{1-leg}+\amp_{2-legs}+\amp_{3-legs}+\amp_{4-legs}+\amp_{res} \,,
\label{amp-by1PI}
\ee
where $\amp^{\rm LO}$ is the result from the LO Lagrangian,  i.e.,  $\amp^{\rm LO}= \amp^{(0)}$,  and  
 $\amp_{n-legs}$ means the contributions to the amplitude from the $n$-legs 1PI functions.  The LO contribution is  that from $\mL_2$ to tree level and therefore it is $\mO(\hbar^0)$.  The  $\amp_{n-legs}$ terms contain the NLO contributions,  including  the contributions from the $a_i$ coefficients and also the $\mO(\hbar^1)$ contributions from the loop diagrams in the corresponding 1PI functions.  Notice also,  that we have separated explicitly the extra contribution to the amplitude from the finite residues of the external particles $\amp_{res}$.  

\begin{figure}[H]
\begin{center}
    \includegraphics[width=.9\textwidth]{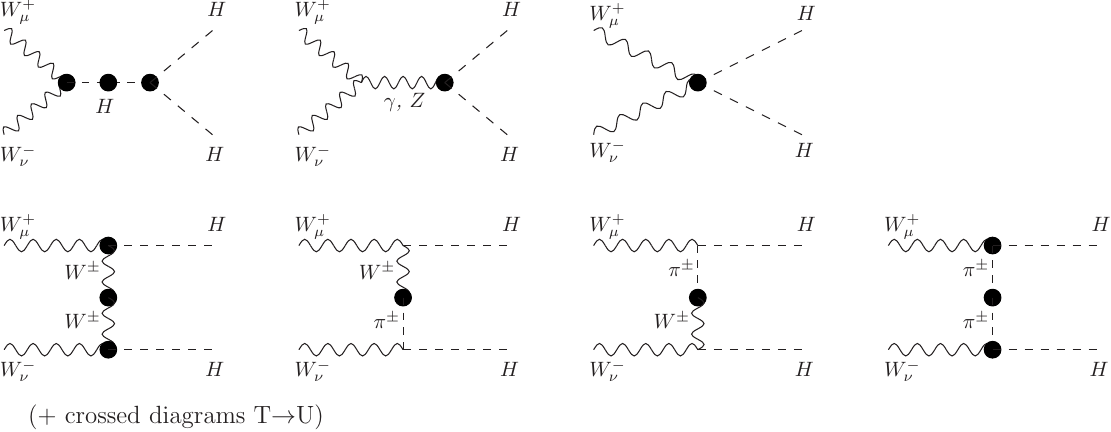} 
\caption{Full 1PI functions (black balls) contributing to the full \1loop  amplitude $ \amp(W^+W^-\to HH)$ }
\label{1PIdiagsWWHHinSTUC}
\end{center}
\end{figure}

From now on, we fix the notation for the momenta assignments and Lorentz indexes for the process of interest as follows: 
\be
W^+_\mu(p_+)\,W^-_\nu(p_-) \to H(k_1)\,H(k_2)\,,
\label{ourscattering}
\ee
where $p_{+,-}$ are the incoming momenta of the gauge bosons, with polarization vectors $\epsilon_+^\mu\equiv\epsilon^\mu(p_+)$ and $\epsilon_-^\nu\equiv\epsilon^\nu(p_-)$, respectively, and $k_{1,2}$ the outgoing momenta of the Higgs bosons (with $p_+ +p_- =k_1 +k_2$). 
Thus,  the amplitude $\amp$ can be written as:
\be
\amp = A_{\mu\nu}\,\epsilon_+^\mu\epsilon_-^\nu \,, 
\label{tensor-amp}
\ee
where the tensor amplitude with explicit Lorentz indexes is defined by $A_{\mu\nu}$. 
 \figref{1PIdiagsWWHHinSTUC} collects the full 1PI functions and full propagators,  represented by black balls, that contribute to the \1loop amplitude 
$\amp(WW\to HH)$. 
These full functions (denoted with a hat) correspond to:  1) the full propagators,  $\propR^{HH}$,   $\propR^{\pi\pi}$, $\propR^{WW}$, $\propR^{W\pi}$, $\propR^{\pi W}$, $\propR^{AA}$ and $\propR^{ZZ}$; 2) the full 1PI vertex functions with three-legs, $\greenfR_{HWW}$, $\greenfR_{\pi WH}$, $\greenfR_{HHH}$, $\greenfR_{AWW}$, $\greenfR_{ZWW}$, $\greenfR_{AHH}$, $\greenfR_{ZHH}$; and 3) the full 1PI vertex function with four-legs, $\greenfR_{WWHH}$. Notice, that some of these full functions receive contributions of both orders, ${\cal O} (\hbar^0)$ and ${\cal O} (\hbar^1)$.  However,  there are some Green functions that vanish at LO and only receive contributions from NLO, such as,  
$\propR^{W\pi}$, $\propR^{\pi W}$, $\greenfR_{AHH}$ and $\greenfR_{ZHH}$. 
This is the reason why the diagrams in \figref{1PIdiagsWWHHinSTUC} involving these particular NLO Green functions have only one black ball, since including two black balls in this case would produce NNLO corrections that are not our aim here.  For the other diagrams,  not involving these particular 1PI functions,  there can appear the product of several black balls containing each one both LO and NLO contributions,  and one has to perform this product accordingly to extract the final result  for the amplitude containing all the terms of both orders,   $\mO(\hbar^0)$ and $\mO(\hbar^1)$. 

The tensor amplitude in \eqref{tensor-amp} is obtained by adding the $s$, $t$, $u$ and contact $c$ channel contributions as represented in \figref{1PIdiagsWWHHinSTUC}.  
These contributions by channels  can be written in terms of the full Green functions as follows:
\bear
i A_s^{\mu\nu} &=& i\greenfR_{HWW}^{\mu\nu}\,i\propR^{HH}\,i\greenfR_{HHH}  \nn\\
&&+i\greenfR_{WWA}^{\mu\nu\rho}\,(-i)\propR_{\rho\sigma}^{AA}\,i\greenfR_{AHH}^{\sigma} + i\greenfR_{WWZ}^{\mu\nu\rho}\,(-i)\propR_{\rho\sigma}^{ZZ}\,i\greenfR_{ZHH}^{\sigma}  \nn\\
i A_t^{\mu\nu} &=& i\greenfR_{HWW}^{\mu\rho}\,(-i)\propR_{\rho\sigma}^{WW}\,i\greenfR_{HWW}^{\sigma\nu} + i\Gamma_{HWW}^{\mu\rho}\,\propR_{\rho}^{W\pi}\,i\Gamma_{\pi WH}^{\nu}  \nn\\
&&+ i\Gamma_{\pi WH}^{\mu}\,\propR_{\sigma}^{\pi W}\,i\Gamma_{HWW}^{\sigma\nu} + i\greenfR_{\pi WH}^{\mu}\,i\propR^{\pi\pi}\,i\greenfR_{\pi WH}^{\nu}  \nn\\
i A_u^{\mu\nu} &=& i\greenfR_{HWW}^{\mu\rho}\,(-i)\propR_{\rho\sigma}^{WW}\,i\greenfR_{HWW}^{\sigma\nu} + i\Gamma_{HWW}^{\mu\rho}\,\propR_{\rho}^{W\pi}\,i\Gamma_{\pi WH}^{\nu}  \nn\\
&&+ i\Gamma_{\pi WH}^{\mu}\,\propR_{\sigma}^{\pi W}\,i\Gamma_{HWW}^{\sigma\nu} + i\greenfR_{\pi WH}^{\mu}\,i\propR^{\pi\pi}\,i\greenfR_{\pi WH}^{\nu}  \nn\\
i A_c^{\mu\nu} &=& i\greenfR_{WWHH}^{\mu\nu} \,.
\label{EChLfullbychannels}
\eear
At \1loop level,  it is convenient to write the full propagators in terms of the self-energies.  
Following our procedure and conventions defined in~\cite{Herrero:2021iqt}, we get the following expressions  for the full propagators in terms of the LO propagators and the full self-energies:  
\bear
i\hat{\Delta}^{HH}(q^2) &=& i\Delta^{HH} + i\Delta^{HH}\,(-i)\SER_{HH}\,i\Delta^{HH} \,,  \nn\\
i\propR^{\pi\pi}(q^2) &=& i\Delta^{\pi\pi} + i\Delta^{\pi\pi}\,(-i)\SER_{\pi\pi}\,i\Delta^{\pi\pi} \,,  \nn\\
-i\propR^{WW}_T(q^2) &=& -i\Delta^{WW}_T -i\Delta^{WW}_T\,i\SER_{WW}^T\,(-i)\Delta^{WW}_T \,,  \nn\\
-i\propR^{WW}_L(q^2) &=& -i\Delta^{WW}_L -i\Delta^{WW}_L\,i\SER_{WW}^L\,(-i)\Delta^{WW}_L \,,  \nn\\
\propR^{W\pi}(q^2) &=& \Delta^{W\pi} -i\Delta^{WW}_L\,\SER_{W\pi}\,i\Delta^{\pi\pi} \,,
\label{oneloop-prop}
\eear
where all functions on the right hand side are functions of $q^2$ and the LO propagators in the $R_\xi$ gauges are summarized by:
\bear
i\Delta^{HH} &=& \frac{i}{q^2 -\mh^2} \,,\qquad -i\Delta^{WW}_T = \frac{-i}{q^2 -\mw^2} \,,  \nn\\
-i\Delta^{WW}_L &=& \frac{-i\xi}{q^2 -\xi\mw^2}  \,,\quad i\Delta^{\pi\pi} = \frac{i}{q^2 -\xi\mw^2} \,,\quad \Delta^{W\pi} = 0 \,.  \nn\\
-i\Delta^{AA}_T &=& \frac{-i}{q^2} \,,\quad -i\Delta^{AA}_L = \frac{-i\xi}{q^2}  \,,\quad -i\Delta^{ZZ}_T = \frac{-i}{q^2 -\mz^2} \,,\quad -i\Delta^{ZZ}_L = \frac{-i\xi}{q^2 -\xi\mz^2}  \,.
\label{propsLO}
\eear
As commented previously, only $\Delta^{HH}$, $\Delta^{WW}_T$, $\Delta^{WW}_L$ and $\Delta^{\pi\pi}$ are involved in the LO contribution to the amplitude in \eqref{EChLfull}.
On the other hand, the relevant vertex functions at LO  are: 
\bear
i\Gamma_{HWW}^{\mu\nu} &=& i a g\mw g^{\mu\nu}  \,,\quad i\Gamma_{HHH}=-3i\kappa_3\mh^2/\vev  \,,  \nn\\
i\Gamma_{\pi WH}^{\mu} &=& ag p_\pi^\mu  \,,\quad i\Gamma_{WWHH}^{\mu\nu}=ibg^2g^{\mu\nu}/2  \,,
\label{vertexLO}
\eear

Next, we present the computation of the LO amplitude using the $R_\xi$ gauges.  This can be easily done by plugging the corresponding LO functions of \eqref{propsLO} and \eqref{vertexLO} in \eqref{EChLfullbychannels}.  Namely, using $\Gamma$ instead of $\hat \Gamma$, and $\Delta$ instead of $\hat \Delta$. 
\begin{figure}[H]
\begin{center}
    \includegraphics[width=.85\textwidth]{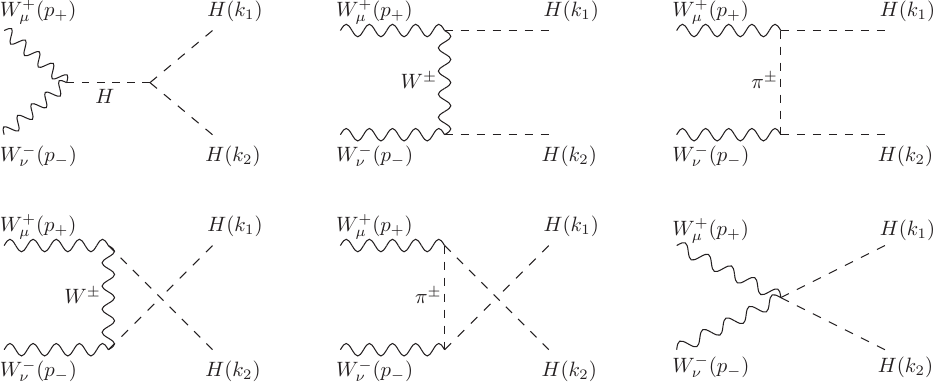}
\caption{Diagrams at LO contributing to $W^+W^-\to HH$ in the covariant $R_\xi$ gauges.}
\label{diagsWWHHatLO}
\end{center}
\end{figure}
 The result for the LO amplitude in the $R_\xi$ gauge,  corresponding to the tree level diagrams in \figref{diagsWWHHatLO} is given by: 
 \bear
 A^{(0)} &=& A^{(0)}_s + A^{(0)}_t + A^{(0)}_u + A^{(0)}_c
 \label{amp-LO}
 \eear 
 where the contributions by $s$,  $t$,  $u$ and contact channels  are given,  respectively,  by:
\bear
A^{(0)}_s &=& \frac{g^2}{2}3a\kappa_3\frac{\mh^2}{S-\mh^2}\epsilon_+\cdot\epsilon_-  \nn\\
A^{(0)}_t &=& g^2a^2\frac{\mw^2\epsilon_+\cdot\epsilon_- +\epsilon_+\cdot k_1\,\epsilon_-\cdot k_2}{T-\mw^2} \,,  \nn\\
A^{(0)}_u &=& g^2a^2\frac{\mw^2\epsilon_+\cdot\epsilon_- +\epsilon_+\cdot k_2\,\epsilon_-\cdot k_1}{U-\mw^2}  \nn\\
A^{(0)}_c&=& \frac{g^2}{2}b\,\epsilon_+\cdot\epsilon_- \,.
\label{amp-LO-bychannels}
\eear
From this equation, notice  that the corresponding result of the SM amplitude at LO is simply obtained from this same formula by setting $a=b=\kappa_3=1$.
We have checked explicitly the gauge invariance of our LO result above,   namely,  that the dependence on $\xi$ disappears in the final amplitude as expected.  The cancellation of the $\xi$-dependent terms is achieved once the external gauge bosons are on-shell, i.e., by contracting the tensorial amplitude with their corresponding polarization vectors in $\amp^{(0)}$.
Concretely, the cancellation of the $\xi$-dependent terms occurs separately in the two channels $t$ and $u$, and it happens between the contribution of the longitudinal part of the $W$ propagator and the GB propagator in \eqref{EChLfullbychannels}.

Finally,  we present the result for the complete amplitude to tree level,  i.e.,  
\bear
\amp^{\treeLtwoLfour}=\amp^{\treeLtwo}+ \amp^{\treeLfour}
\label{amp-tree}
\eear
where $\amp^{\treeLtwo}= A^{(0)}$ is given in \eqrefs{amp-LO}{amp-LO-bychannels} and $\amp^{\treeLfour}$ is computed from $\mL_4$ and contains the $a_i$ coefficients.  As we have explained in the previous section,  this can be written in two ways,  depending if one uses the EOMs to reduce the list of operators or not.  We provide here the short version, i.e.,  using ${\mL}_{4}^{\rm +EOMs}$ in \eqref{eq-L4-con-eoms-completa}. 
\bear
\amp^{\treeLfour}&=& \amp^{\treeLfour}\vert_s +  \amp^{\treeLfour}\vert_t + \amp^{\treeLfour}\vert_u+ \amp^{\treeLfour}\vert_c
\eear
where the contributions by channels are:
\bear
\amp^{\treeLfour}\vert_s &=& \frac{g^2}{2\vev^2}\frac{1}{S-\mh^2}\left(3\kappa_3a_{d2}\mh^2(S\epsilon_+\cdot\epsilon_- -2\epsilon_+\cdot p_-\,\epsilon_-\cdot p_+)  \right.  \nn\\
&&\left. \hspace{3mm}+6\kappa_3a_{HWW}\mh^2((S-2\mw^2)\epsilon_+\cdot\epsilon_- -2\epsilon_+\cdot p_-\,\epsilon_-\cdot p_+)  \right.  \nn\\
&&\left. \hspace{3mm}-(3\kappa_3\tilde{a}_{H\mV\mV}\mh^4+a(\tilde{a}_{Hdd}\mh^2+a_{ddW}\mw^2+a_{ddZ}\mz^2)(S+2\mh^2))\epsilon_+\cdot\epsilon_- \right)  \nn\\
\amp^{\treeLfour}\vert_{t} &=& \frac{g^2}{2\vev^2}\frac{a}{T-\mw^2} \left(a_{d2}(4\mw^2\mh^2\epsilon_+\cdot\epsilon_- +2(T+3\mw^2-\mh^2)\epsilon_+\cdot k_1 \epsilon_-\cdot k_2  \right.  \nn\\
&& \left. \hspace{30mm}-4\mw^2(\epsilon_+\cdot k_1\epsilon_-\cdot p_+ +\epsilon_+\cdot p_-\epsilon_-\cdot k_2 ))  \right.  \nn\\
&& \left. \hspace{12mm}-8a_{HWW}\mw^4((T+\mw^2-\mh^2)\epsilon_+\cdot\epsilon_- +\epsilon_+\cdot k_1 \, \epsilon_-\cdot p_+ +\epsilon_+\cdot p_- \, \epsilon_-\cdot k_2)  \right.  \nn\\
&& \left. \hspace{12mm}-4\tilde{a}_{H\mV\mV}\mh^2(\mw^2\epsilon_{+}\cdot\epsilon_{-} +\epsilon_{+}\cdot k_1\,\epsilon_{-}\cdot k_2)  \right)  \nn\\  
\amp^{\treeLfour}\vert_{u} &=& \amp^{\treeLfour}\vert_{t} \,\,\,\rm{with}\,\,\,T\to U \,\,\,\rm{and}\,\,\, k_1\leftrightarrow k_2  \nn\\
\amp^{\treeLfour}\vert_c &=& \frac{g^2}{2\vev^2}\left(-2 \tilde{a}_{dd\mV\mV1} (\epsilon_+\cdot k_2 \, \epsilon_-\cdot k_1+\epsilon_+\cdot k_1 \, \epsilon_-\cdot k_2)  \right.\nn\\
&&\left. \hspace{3mm}+(-2\tilde{a}_{dd\mV\mV2}(S-2\mh^2) +4a_{HHWW}(S-2\mw^2) +a_{Hd2}S -\tilde{a}_{HH\mV\mV}\mh^2)\epsilon_+\cdot\epsilon_-  \right.\nn\\
&&\left. \hspace{3mm}-2 (a_{Hd2}+4 a_{HHWW}) \epsilon_+\cdot p_- \, \epsilon_-\cdot p_+ \right)
\label{ampWWtoHH-HEFT}
\eear
Notice that we have used the new coefficients defined in \eqref{delta-eta}.  Notice also that the above results are given in terms of the polarization vectors of the initial $W$ gauge bosons. Therefore, our results above apply to all the possible polarized channels, $W_X W_Y \to HH$, with $XY=LL, TT,LT,TL$ by just inserting the proper polarization vectors $\epsilon_+$ and $\epsilon_-$. 

We next comment shortly on the comparison of our analytical results in this section for the tree level amplitude within  the EChL with the previous literature.  First of all,  the LO amplitude  in \eqrefs{amp-LO}{amp-LO-bychannels} is in full agreement with 
\cite{Gonzalez-Lopez:2020lpd}.  Secondly,  regarding  $ \amp^{\treeLfour}$ we have compared our results with those in \cite{Asiain:2021lch}. We have checked the full agreement in the contributions from the coefficients,  $\tilde{a}_{dd\mV\mV1}$ ($=\eta$),  
$\tilde{a}_{dd\mV\mV2}$ ($=\delta$),  $a_{d2}$ ($=b_1 \chi$) and $a_{Hd2}$ ($=2b_2 \chi$) with their results.  The other coefficients in our result of 
\eqref{ampWWtoHH-HEFT} were not considered in \cite{Asiain:2021lch}.  On the other hand,  the results of \cite{Delgado:2013hxa} in terms of $\delta$ and $\eta$ where provided using the Equivalence Theorem,  so they can only be compared for the longitudinal modes and in the high energy regime $\sqrt{s}\gg \mw, \mh$.   By an exploration of our amplitudes for the case of the longitudinal modes  in that high energy regime,  we have also checked agreement of the  $\eta$ and $\delta$ contributions with that reference.  The other parameters were not studied either in that reference.

\begin{figure}[ht!]
	\begin{center}			\includegraphics[scale=0.5]{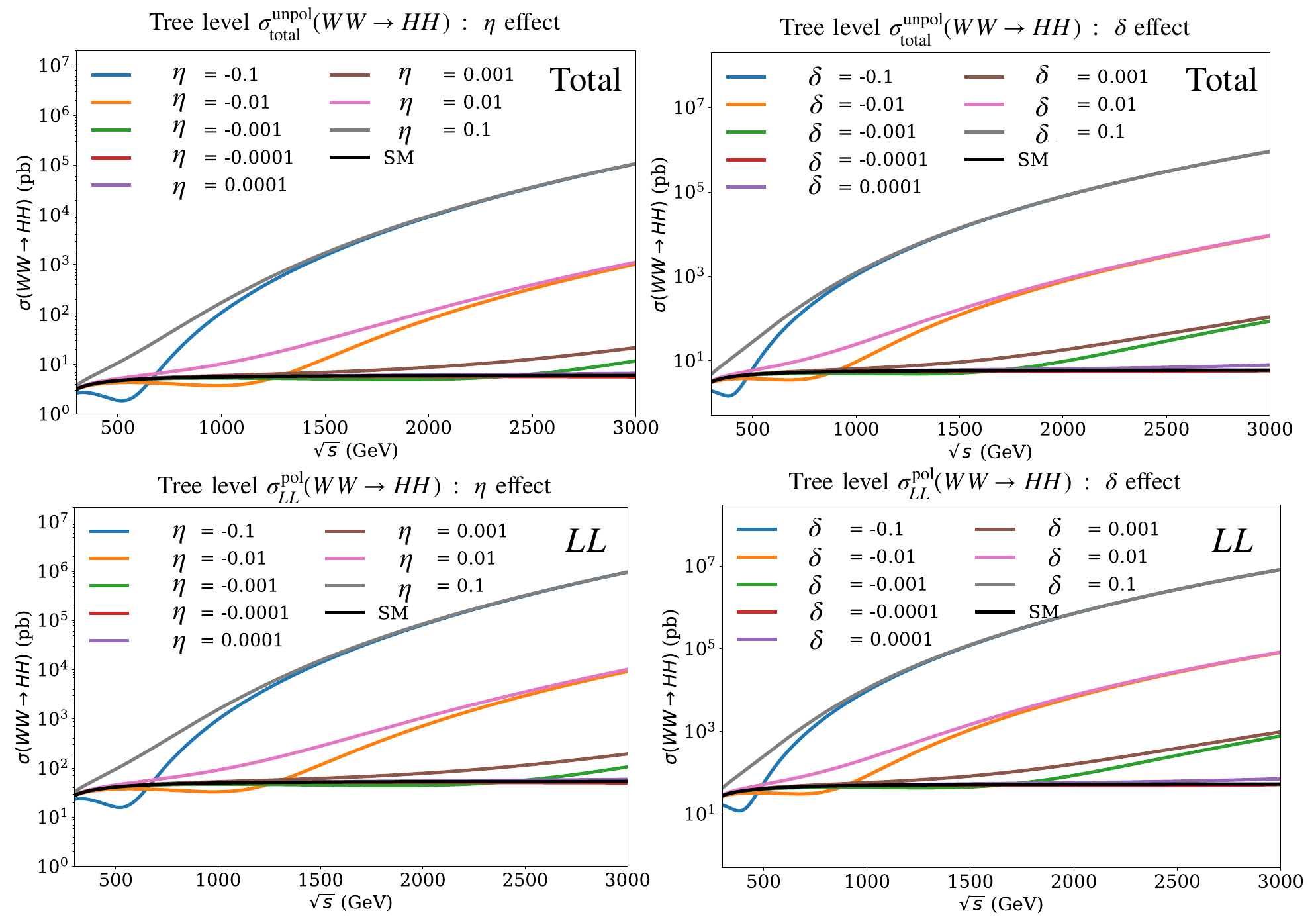} 
	\caption{Tree level cross section predictions for $W^+ W^-\to HH$ within the EChL setting $a=b=\kappa_3=\kappa_4=1$. All EChL coefficients in the NLO Lagrangian are set to zero except for $\eta$ and $\delta$.  Plots in the left column are for non-vanishing $\eta$ and plots in the right column are for non-vanishing $\delta$.  The predictions for the total unpolarized case are displayed in the plots of the first row, and the ones for the polarized $LL$ case in the second row.  The SM predictions are displayed in all the plots, for comparison. }
	\label{plot-xsTree}
	\end{center}
\end{figure}

Finally, it is important to keep in mind the existent hierarchy among the various polarization channels and among the relevance of the various coefficients for each polarization channel.  Firstly, it is well known the dominance in the total cross section for this $WW\to HH$ process  of the longitudinal polarized  modes over the transverse modes.  Namely,  $\sigma(WW \to HH)$ is fully dominated by $\sigma(W_LW_L \to HH)$. The other polarization channels with initial $W_T W_T$ or $W_LW_T$ are highly subdominant at the center of mass energies in the TeV domain. Therefore,  by studying the longitudinal polarized case one can approximate quite well the total cross section.  This dominance of the $\sigma (W_LW_L \to HH)$ over the other polarized channels also happens in the EChL case,  in the tree level estimates of the cross section,  at both orders the LO and the NLO ones.  A recent phenomenological study of the corresponding BSM effects in Ref. \cite{RoberMariaDaniMJ} for  all the polarized channels and considering all the EChL coefficients in \eqref{ampWWtoHH-HEFT},  has shown that the most relevant coefficients of the EChL,  for the $LL$ modes and  at the tree-level NLO,  are indeed $\eta$ and $\delta$.  Here,  by `the most relevant coefficients' we mean those EChL coefficients in  $\mL_4$ that lead to the largest cross sections in this $WW \to HH$ scattering process at the TeV energy domain.   For definiteness here,  and to summarize this $LL$ dominance in the tree level-NLO  prediction from the EChL,  we show in \figref{plot-xsTree}  our predictions,  as a function of the center-of-mass-energy $\sqrt{s}$,  of the cross sections: 1)  for the total unpolarized case (the two plots on the first row),  and 2) for the $LL$ polarized case  (the two plots on the second row).  We display in this figure the BSM departures  respect to the SM predictions from the separate effects of the two most relevant coefficients, assuming different numerical values for those coefficients ($\pm 0.1, 0.01, 0.001, 0.0001$): 1) the effect from $\eta$ is displayed in the plots on the first column; 2) the effect from  $\delta$ is displayed in the plots on the second column.  We can clearly see in these plots that the cross section 
for the $LL$ case fully dominates the total (unpolarized) cross section for all the studied cases.  Indeed, the two lines for $LL$ and for  `total' practically coincide in the studied TeV domain (up to the obvious reducing 1/9 factor in the unpolaized result due to the average over the possible initial helicities).  The other evident conclusion from this figure is that large values of the cross sections and large departures from the SM predictions can be reached at the TeV energies for the cases with the larger input coefficients $\eta$ and $\delta$.  For a more devoted study of the phenomenological consequences of these tree level predictions within the EChL at NLO we address the reader to  Ref. \cite{RoberMariaDaniMJ}.  In particular,  the relevance of these predictions for the di-Higgs production at future $e^+e^-$ colliders via $WW$ fusion has also been explored in that reference.  In the following part of the present work,  we do not go further in these phenomenological issues and focus instead in our main purpose here: the computation of the EW radiative corrections for the $WW \to HH$ scattering process. 

\section{Renormalization procedure}
\label{sec-renorm}
\subsection{Generalities}
In this section we present our renormalization program to compute the renormalized 1PI functions within the EChL in covariant $R_\xi$ gauges using a diagrammatic approach.
These renormalized 1PI functions, denoted here generically by $\greenfR$, receive contributions from the tree level Lagrangian $\mL_2 +\mL_4$, $\greenfT$; from the \1loop diagrams using the interaction vertices of $\mL_2$ only, $\greenfL$;  and from all the counterterms of $\mL_2 +\mL_4$, $\greenfC$:
\be
\greenfR_{n-legs} =\greenfT_{n-legs}+\greenfL_{n-legs}+\greenfC_{n-legs} \,.
\label{fgreen}
\ee
Notice again the double role of ${\cal L}_4$ in the chiral Lagrangian approach: on the one hand, it contributes to a tree level scattering amplitude, and on the other hand it also acts as source of new counterterms in order to remove the extra divergences emerging from the loops computed with $\mL_2$,  which are not removable by a simple redefinition of the parameters in this part of the Lagrangian.

Our analytical computation here is performed with the various softwares associated to Wolfram Mathematica~\cite{WMathematica} and starts by implementing our model in FeynRules~\cite{FeynRules}, generating and drawing the Feynman diagrams with FeynArts~\cite{FeynArts} and performing the main calculations with FormCalc and LoopTools~\cite{FormCalc-LT}. Some extra checks of the involved \1loop divergences were made using FeynCalc~\cite{FeynCalc} and Package-X~\cite{packX}.
The SM results were obtained following the same steps.

The renormalization program followed in this work is similar to the one we already presented in~\cite{Herrero:2021iqt} in the EChL context for Vector Boson Scattering (VBS) processes,  like $WZ \to WZ$, etc. 
Next we briefly summarize the main aspects of the regularization and multiplicative renormalization prescriptions, the renormalization conditions and then we present the new one-loop diagrams,  the new divergences,  and  the solutions for all the counterterms relevant for $WW\to HH$ scattering.

\subsection{Regularization and Renormalization prescriptions}
\label{RR}
As it is usual, our regularization procedure of the loop contributions is performed with dimensional regularization~\cite{Bollini:1972ui,tHooft:1972tcz} in $D=4-\epsilon$ dimensions. 
This method preserves all the relevant symmetries in the bosonic sector of the theory, including chiral invariance (Dirac $\gamma_5$ is not involved in this work since we do not consider the fermionic contributions).  
Consequently,  the scale of dimensional regularization is set to $\mu$ and all the \1loop divergences are expressed in terms of:
 \be
\div=\frac{2}{\epsilon}-\gamma_E+\log(4\pi) \,.
\label{div-definition}
\ee
Concerning the renormalization procedure, we generate the counterterms of all the parameters and fields appearing in the tree level Lagrangian, $\mL_2+\mL_4$, by the usual multiplicative renormalization prescription that relates the bare quantities (here denoted by a specific sub- or super-script with a label $0$) and the renormalized ones (here with no specific sub- or super-script labels).  
We have the following relations:
\bear
H_0 &= & \sqrt{\Zf_H}H \,, \quad B_{0\,\mu} = \sqrt{\Zf_B}B_{\mu}\,, \quad W_{0\,\mu}^{1,2,3} = \sqrt{\Zf_W}W_{\mu}^{1,2,3} \,, \quad \pi_0^{1,2,3} = \sqrt{\Zf_\pi}\pi^{1,2,3} \,,  \nn\\
\vev_0 &= & \sqrt{\Zf_\pi} (\vev +\delta\vev)\,,
\quad \lambda_0 = \Zf_H^{-2}(\lambda +\delta \lambda)\,, \nn\\
\gY_0 &= & \Zf_B^{-1/2}(g' +\delta \gY)\,,\quad
g_0 = \Zf_W^{-1/2}(g +\delta g)\,, \quad \xi_{1,2}^0 = \xi (1+\delta\xi_{1,2})\,,  \nn\\
a^0&=&a+\delta a\,,\quad b^0=b+\delta b\,,\quad \kappa_{3,4}^0=\kappa_{3,4}+\delta\kappa_{3,4}\,,\quad a_i^0=a_i+\delta a_i\,,
\label{EChL-renorm-factors}
\eear
where the $Z_i=1+\delta\Zf_i$ are the usual renormalization multiplicative constants and we use the generic notation $\delta p$ ($\delta a_i$) for the counterterm of each involved EW parameter $p$ (effective coefficient $a_i$).

With these definitions, our final results for both the renormalized 1PI functions and the $WW\to HH$ scattering amplitude are expressed in terms of the renormalized quantities, $\mw$, $\mz$, $\mh$, $g$, $\gY$, $\vev$,  $a$, $b$, $\kappa_3$, $\lambda$ and the $a_i$'s.  
Notice that $\kappa_4$ and the ghost counterterms do not enter in the present computation and we omit to show them for shortness.  
On the other hand, the renormalization of the covariant gauge fixing parameters have set to a common renormalized $\xi$ parameter for all the involved EW gauge bosons.  For more details on the technicalities of our renormalization method,  see Ref. \cite{Herrero:2021iqt}.  

Next, we  summarize the renormalization conditions.  As in Ref. \cite{Herrero:2021iqt} we adopt here a hybrid prescription in which we choose the on-shell (OS) scheme for the EW parameters in the lowest order Lagrangian $\mL_2$ and the $\overline{MS}$ scheme for all the EChL coefficients.  The list of conditions are as follows:
\begin{itemize}
    \item Vanishing (Higgs) tadpole:
        \be
        \hat{T}=0 \,.
        \label{condOS1}
        \ee
    \item The pole of the renormalized propagator of the Higgs boson lies at $\mh^2$ and the corresponding residue is equal to 1:
        \be
        {\rm Re}\left[ \SER_{HH}(\mh^{2}) \right] =0 \,,\quad {\rm Re}\left[ \frac{d\SER_{HH}}{dq^2}(\mh^{2}) \right] =0 \,.
        \label{condOS2}
        \ee
    \item Properties of the photon: residue equal one; no $A-Z$ mixing propagators; and the electric charge defined like in QED, since there is a remnant $U(1)_{\rm em}$ electromagnetic gauge symmetry:
        \be
        {\rm Re}\left[ \frac{d\SER^T_{AA}}{dq^2}(0) \right] =0 \,,\quad \SER^T_{ZA}(0) =0  \,,\quad \greenfR^\mu_{\gamma e e}\vert_{\rm OS}=i e \gamma^\mu \,. 
        \label{condOS3}
        \ee    
    \item The poles of the transverse renormalized propagators of the $W$ and $Z$ bosons lie at $q^2=\mw^2$ and $q^2=\mz^2$,  respectively:
        \be
        {\rm Re}\left[ \SER_{WW}^T(\mw^{2}) \right] =0 \,,\quad {\rm Re}\left[ \SER_{ZZ}^T(\mz^{2}) \right] =0 \,.
        \label{condOS4}
        \ee
 \item The poles of the renormalized propagators in the unphysical charged sector $\{W^\pm,\pi^\pm\}$ lie at $q^2=\xi\mw^2$. Therefore:
        \be
        {\rm Re}\left[ \SER_{WW}^L(\xi\mw^{2}) \right] =0 \,,\quad {\rm Re}\left[ \SER_{\pi\pi}(\xi\mw^{2}) \right] =0 \,.  
        \label{condOS5}
        \ee
    \item $\overline{MS}$ scheme for all the involved EChL coefficients. \\
    In particular for $a$,  $b$,  $\kappa_{3}$,  $\kappa_{4}$  in \eqref{eq-L2} and the $a_i$'s  in \eqref{eq-L4-without-eoms}
\end{itemize}

The above renormalization conditions on all the EChL parameters determine both the divergent and finite parts involved in all the 1PI functions,  and therefore also in the one-loop scattering amplitudes.  Notice that the residue for the Higgs and photon fields are set to one in the previous conditions, but the resulting residues ${\cal Z}_{W(Z)}$ of the gauge bosons $W$($Z$), are different to one. 
Since each external $W$ provides a factor ${\cal Z}^{1/2}_{W}$ to the observable $S$ matrix, then the corresponding contribution from the residues ( $\amp_{res}$ in \eqref{amp-by1PI}) of the two external $W$'s in  the  $WW\to HH$ scattering is given by:
\be
\amp_{res} = {\rm Re}\left[ \frac{d\SER^T_{WW}}{dq^2}(\mw^2) \right] \amp^{(0)} \,,
\label{ampRes}
\ee
In addition, the Higgs tadpole enters in many parts of the different diagrams contributing to the amplitude. However, with the renormalization condition of \eqref{condOS1}, the $\amp_{1-leg}$ in \eqref{amp-by1PI} vanishes.

\subsection{Summary of contributions to the renormalized 1PI functions}
\label{1PIR}

We emphasize again that our renormalization program in the $R_\xi$ gauges makes finite all the relevant 1PI Green functions for arbitrary momentum of the external legs (hence, generically, off-shell) and no transversality condition for the EW gauge bosons, $p_i\cdot\epsilon(p_i)=0$, is applied,  for those 1PI results.  This means,  that our renormalization program is  more demanding than the usual renormalization program which gets just finite results for the scattering amplitudes with on-shell external legs.  Notice also that in this later case the trasversality conditions for the external gauge bosons are usually used as well.

In the following of this section, we collect the various contributions to the renormalized 1PI functions, already mentioned in \secref{diag-1pi}, that enter in $WW\to HH$ scattering and that were not involved in our previous computation~\cite{Herrero:2021iqt} which was addressed to the $WZ \to WZ$ case.  In particular,  we exhibit now the results for the Green functions involving Higgs bosons in the external legs, corresponding to the vertices $HHH$, $HWW$, $\pi WH$, $AHH$, $ZHH$ and $WWHH$.  And,  for completeness, we also include 
 in \appref{App-previous-results} a short summary of the other renormalized 1PI functions derived in~\cite{Herrero:2021iqt} that also enter here,  for the $WW \to HH$  scattering.  For definiteness, all the explicit analytical results presented in the present paper (and in the appendices) are provided in the Feynman 't Hooft gauge  with $\xi=1$.  

The results of the 3-legs functions corresponding to $H(p_1)H(p_2)H(p_3)$, $H(q)W^\mu(k_1)W^\nu(k_2)$, $\pi(q)W^\mu(p_W)H(p_H)$ and $V^\rho(q)H(p_1)H(p_2)$ (with $V=A,\,Z$),  are given by the sum of the LO part (if any),  loop contributions,  EFT coefficient's contributions and CT contributions,   as follows:
\bear
i\greenfR_{HHH} &=& -3i\kappa_3\frac{\mh^2}{\vev} +i\greenfL_{HHH} -3i\kappa_3\frac{\mh^2}{\vev}\left(\frac{\delta\kappa_3}{\kappa_3}+\frac{\delta\mh^2}{\mh^2}-\frac{\delta\Zf_\pi}{2}-\frac{\delta\vev}{\vev}+\frac{3\delta\Zf_H}{2}\right)  \nn\\
&& +\frac{i}{\vev^3}\left(a_{dd\Box}(p_1^4+p_2^4+p_3^4)+2(a_{H\Box\Box}-a_{dd\Box})(p_1^2p_2^2+p_2^2p_3^2+p_3^2p_1^2) \right.\nn\\
&&\left. \hspace{10mm}+(a_{Hdd}\mh^2+a_{ddW}\mw^2+a_{ddZ}\mz^2)(p_1^2+p_2^2+p_3^2)\right)  \nn\\
i\greenfR_{HW^-W^+}^{\mu\nu} &=& ia\frac{g^2\vev}{2}\gmunu +i\greenfL_{HWW} +i\frac{ a g^2 \vev}{2}\left(\frac{\delta a}{a}+\frac{2 \delta g}{g}+\frac{\delta\vev}{\vev}+\frac{\delta\Zf_H}{2}+\frac{\delta\Zf_\pi}{2}\right)\gmunu  \nn\\
&& -i\frac{g^2}{2\vev}\left(\left(-(2a_{HWW}+a_{d2}+2a_{\Box{\cal V}{\cal V}})q^2+2a_{HWW}(k_1^2+k_2^2) +a_{H\mV\mV}\mh^2\right) \gmunu \right.  \nn\\
&&\left.\hspace{12mm}+\left(a_{d2}+a_{d3}\right)\left(k_1^\mu k_1^\nu+k_2^\mu k_2^\nu\right)+2\left(a_{d3}-a_{H11}\right)k_1^\mu k_2^\nu+2\left(2a_{HWW}+a_{d2}\right)k_2^\mu k_1^\nu \right) \,,  \nn\\
i\greenfR_{\pi^+ W^-H}^{\mu} &=& -iag(p_W+p_H)^\mu +i\greenfL_{\pi WH} -iag\left(\frac{\delta a}{a}+\frac{\delta g}{g}+\delta\Zf_H/2+\delta\Zf_\pi/2\right)(p_W+p_H)^\mu \nn\\
&& +i\frac{g}{2\vev^2}\left(-2 (p_H^\mu+p_W^\mu) (2 a_{\Box {\cal V}{\cal V}} p_H^2+a_{d2} p_W^2-a_{d3} q^2 -a_{H\mV\mV}\mh^2)\right. \nn\\
&&\left.\hspace{13mm}+p_W^\mu q^2 (a_{d2}-3 a_{d3}+4 a_{H11})+p_W^\mu (a_{d2}+a_{d3}) (p_W^2-p_H^2)\right)  \nn\\
i\greenfR_{AHH}^{\rho} &=& i\greenfL_{AHH}  \,,  \nn\\
i\greenfR_{ZHH}^{\rho} &=& i\greenfL_{ZHH} \,.
\label{new3legs-renorm}
\eear
Similarly,  the result of the 4-legs function corresponding to $W^\mu(p_1)W^\nu(p_2)H(p_3)H(p_4)$ is given by the sum of the LO part, the loop contributions, the EFT coefficients contributions  and the CT contributions  as follows, 
\bear
i\greenfR_{W^+W^-HH}^{\mu\nu} &=& i\frac{bg^2}{2}\gmunu +i\greenfL_{WWHH} +i\frac{bg^2}{2}\left(\frac{\delta b}{b}+\frac{2\delta g}{g}+\delta\Zf_H\right)\gmunu  \nn\\
&&\hspace{-8mm} -\frac{ig^2}{2\vev^2}\left(\gmunu(-(p_3^2+p_4^2)(-2a_{dd\mV\mV2}+4a_{HHWW}+2a_{H\Box{\cal V}{\cal V}}) +(p_1+p_2)^2(-2a_{dd\mV\mV2}-a_{Hd2})\right. \nn\\
&&\left.\hspace{15mm} +4((p_1+p_3)^2+(p_1+p_4)^2)a_{HHWW} +a_{HH\mV\mV}\mh^2)\right. \nn\\
&&\left.\hspace{10mm} +4(a_{HH11}-2a_{HHWW})p_2^\mu p_2^\nu -(a_{Hd2}-a_{Hd3}+8a_{HHWW})p_2^\mu(p_3^\nu+p_4^\nu) \right. \nn\\
&&\left.\hspace{10mm} +(a_{Hd2}-a_{Hd3}+4a_{HHWW})(p_3^\mu+p_4^\mu)p_2^\nu +(a_{Hd2}+a_{Hd3})(p_3^\mu p_3^\nu+p_4^\mu p_4^\nu) \right. \nn\\
&&\left.\hspace{10mm} +(-2a_{dd\mV\mV1}+a_{Hd2}+a_{Hd3})(p_3^\mu p_4^\nu+p_4^\mu p_3^\nu)\right)
\label{WWHH-renorm}
\eear
In all the previous expressions above,  in \eqrefs{new3legs-renorm}{WWHH-renorm},  the explicit $a_i$ coefficients entering are  the bare $a_i^0$ coefficients, but for shortness we have dropped the superindex $0$.  Therefore,  the $a_i$'s included  in these equations must be all understood rather  as $(a_i+\delta a_i)$,  with these $a_i$'s being the renormalized coefficients in the $\overline{MS}$,  and $\delta a_i$ the corresponding divergent CT needed to cancel the new divergences from the loops of the 1PI functions. 
As we have said,  the computation of the loop contributions to all these 1PI functions are performed with the help of FormCalc and LoopTools.  For illustrative purposes, we show in \figrefs{new3legs-generic-loops}{WWHH-generic-loops},  all the generic one-loop diagrams entering in the computation of the previous $ \Gamma^{\rm loop}$ functions.  Notice, that since we are working with covariant $R_\xi$ gauges we have considered all the possible particles propagating in the loops, namely, GBs, Higgs,  EW gauge bosons and ghosts.  

Next we provide our results for the divergent (singular) parts of these Loop contributions for the relevant 1PI functions in \eqrefs{new3legs-renorm}{WWHH-renorm}.
All these divergent contributions will set the values of the ${\cal O}(\Delta_\epsilon)$ counterterms, both for the EW parameters and the $a_i$ coefficients, that are relevant for our computation. We get the following results:
\bear
i\greenfL_{HHH}\vert_{div} &=& i\frac{\Delta_\epsilon}{16\pi^2}\frac{3}{2\vev^3}\left(9\kappa_3\kappa_4\mh^4+12ab(2\mw^4+\mz^4) -2a^3(p_1^2p_2^2+p_2^2p_3^2+p_3^2p_1^2)\right. \nn\\
&&\left.\hspace{10mm}-a(a^2-b)(p_1^4+p_2^4+p_3^4-2(2\mw^2+\mz^2)(p_1^2+p_2^2+p_3^2)) \right)  \nn\\
i\greenfL_{HWW}\vert_{div} &=& i\frac{\Delta_\epsilon}{16\pi^2} \frac{g^2}{12\vev} \left(\left(3a(2+a^2)q^2 +a(a^2-b)(k_1^2+k_2^2)  \right.\right. \nn\\
&&\left.\left.\hspace{20mm}-3(a^2-b)(2a-3\kappa_3)\mh^2 -18ab\mw^2+78a\mw^2-18a \mz^2\right)\gmunu \right.  \nn\\
&&\left.\hspace{18mm} +2a(a^2+2b)(\kTmu\kVnu+\kVmu\kTnu) +12a^3\kTmu\kTnu \right)  \nn\\
i\greenfL_{\pi WH}\vert_{div} &=& i\frac{\Delta_\epsilon}{16\pi^2}\frac{g}{6\vev^2}\left((p_H^\mu+p_W^\mu)(-9(a^2-b)\kappa_3\mh^2 -a^3(3p_H^2+p_W^2-6\mh^2+3q^2) \right. \nn\\
&&\left. \hspace{18mm}+a(-6p_H^2-34\mw^2+14\mz^2+b(p_W^2-6\mh^2+18\mw^2-3q^2))) \right. \nn\\
&& \left. \hspace{10mm}p_W^\mu a((2+a^2+2b)p_H^2-a^2(p_W^2-11q^2)+2(\mw^2+\mz^2-b(p_W^2+q^2)))\right)  \nn\\
i\greenfL_{AHH}\vert_{div} &=& i\greenfL_{ZHH}\vert_{div} =0  \nn\\
i\greenfL_{WWHH}\vert_{div} &=& -i\frac{\Delta_\epsilon}{16\pi^2}\frac{g^2}{12\vev^2}\left( \gmunu(3(-8a^4+12a^3\kappa_3-12ab\kappa_3+a^2(10b-3\kappa_4)-2b^2+3b\kappa_4)\mh^2\right. \nn\\
&&\left.\hspace{20mm} +3b((6b-26)\mw^2+6\mz^2) \right. \nn\\
&&\left.\hspace{15mm}(p_3^2+p_4^2)a^2(6+6a^2-3b)\right. \nn\\
&&\left.\hspace{13mm}+6(p_1+p_2)^2(1+a^2)(a^2-b)+((p_1+p_3)^2+(p_1+p_4)^2)(4a^4-5a^2b+b^2))\right. \nn\\
&&\left.\hspace{10mm} -8(4a^4-5a^2b+b^2)p_2^\mu p_2^\nu +2(4a^4+a^2b-2b^2)p_2^\mu(p_3^\nu+p_4^\nu) \right. \nn\\
&&\left.\hspace{10mm} -2(20a^4-19a^2b+2b^2)(p_3^\mu+p_4^\mu)p_2^\nu +2(4a^4+a^2b-2b^2)(p_3^\mu p_3^\nu+p_4^\mu p_4^\nu) \right. \nn\\
&&\left.\hspace{10mm} +6a^2(2a^2+b)(p_3^\mu p_4^\nu+p_4^\mu p_3^\nu) \right)
\label{div-Loop}
\eear

Finally, we present the corresponding results in the SM for the Green functions that are involved in the $WW\to HH$ computation and were not given in~\cite{Herrero:2021iqt}. We use the `bar' notation for all the 1PI  functions in the SM, not to be confused with the previous ones of the HEFT.  Notice  that, contrary to the HEFT,  in the SM case, the multiplicative renormalization constant for the Higgs and GBs fields are the same ($\Zf_\phi$) since they form a doublet.  We get the following SM results:
\bear
i\SMgreenfR_{HHH} &=& -3i\frac{\mh^2}{\vev} +i\SMgreenfL_{HHH} -3i\frac{\mh^2}{\vev}\left(\frac{\delta\mh^2}{\mh^2}+\delta\Zf_\phi-\frac{\delta\vev}{\vev}\right)  \nn\\
i\SMgreenfL_{HHH}\vert_{div} &=& i\frac{\Delta_\epsilon}{16\pi^2}\frac{3}{\vev^3}\left(6\mh^4+6(2\mw^4+\mz^4)-\mh^2(2\mw^2+\mz^2)\right)  \nn\\
i\hat{\overline{\Gamma}}_{HW^+W^-}^{\mu\nu} &=& i\frac{g^2\vev}{2}\gmunu +i\SMgreenfL_{HWW} +i\frac{ g^2 \vev}{2}\left(\frac{2 \delta g}{g}+\frac{\delta\vev}{\vev}+\delta\Zf_\phi\right)\gmunu \,,  \nn\\
i\SMgreenfL_{HWW}\vert_{div} &=& i\frac{\Delta_\epsilon}{16\pi^2} \frac{g^2}{2\vev} \left(10\mw^2-3\mz^2 \right)\gmunu  \nn\\
i\SMgreenfR_{\pi WH}^\mu &=& -i\frac{g}{2}(p_W+2p_H)^\mu +i\SMgreenfL_{\pi WH} -i\frac{g}{2}\left(\frac{\delta g}{g}+\delta\Zf_\phi\right)(p_W+2p_H)^\mu  \nn\\
i\SMgreenfL_{HW\pi}\vert_{div} &=& -i\frac{\Delta_\epsilon}{16\pi^2}\frac{g}{\vev^2}(2\mw^2-\mz^2)(p_W+2p_H)^\mu  \nn\\
i\SMgreenfR_{WWHH}^{\mu\nu} &=& i\frac{g^2}{2}\gmunu +i\SMgreenfL_{WWHH} +i\frac{g^2}{2}\left(\frac{2\delta g}{g}+\delta\Zf_\phi\right)\gmunu  \nn\\
i\SMgreenfL_{WWHH}\vert_{div} &=& i\frac{\Delta_\epsilon}{16\pi^2}\frac{g^2}{\vev^2}\left(6\mw^2-\mz^2\right)\gmunu  \nn\\
i\SMgreenfR_{AHH}^{\rho} &=& i\SMgreenfL_{AHH}  \nn\\
i\SMgreenfL_{AHH}\vert_{div} &=& 0  \nn\\
i\SMgreenfR_{ZHH}^{\rho} &=& i\SMgreenfL_{ZHH}  \nn\\
i\SMgreenfL_{ZHH}\vert_{div} &=& 0
\label{SM1PIfunctions}
\eear

\subsection{Renormalization of the EFT parameters}
\label{sec-main-div-results}

In this section we present the results for the renormalization of the EFT parameters.  These include the EW parameters entering in $\mL_2$,  like $g,$  $g'$,  etc., and the EChL coefficients,  namely,  $a$,  $b$, $\kappa_3$ entering in $ \mL_2$,  and the $a_i$ coefficients entering in $ \mL_4$. 

First, we determine the divergent parts (called in short $\delta_\epsilon$) of all the counterterms requiring that all the renormalized 1PI functions at arbitrary values of the external leg momenta (generically off-shell) results finite.
This procedure leads to a system of equations by demanding the cancellation of the ${\cal O}(\Delta_\epsilon)$ contributions for each involved Lorentz structure and in each term in the momentum powers expansion of the Green functions, that it is solved sequentially.
The CTs corresponding to the $\mL_2$ parameters in \eqref{eq-L2}, except for $b$, $\kappa_3$ and $\lambda$, and some of the $a_i$'s coefficients in \eqref{eq-L4-without-eoms} were already derived in our previous work~\cite{Herrero:2021iqt}.
Respect to this reference, we add now the Green functions with Higgs bosons corresponding to the vertices  $HHH$, $HWW$, $\pi WH$ and $WWHH$ (notice that the corresponding ones to $AHH$ and $ZHH$ are finite and do not have new EChL coefficients). 
In particular, we derive $\deltaCT\lambda$ from the tadpole's counterterm; $\hat{\Gamma}_{HHH}$ sets $\deltaCT\kappa_3$, $\deltaCT a_{dd\Box}$, $\deltaCT a_{H\Box\Box}$ and $\deltaCT a_{Hdd}$; $\hat{\Gamma}_{HWW}$ sets $\deltaCT a_{HWW}$, $\deltaCT a_{d2}$, $\deltaCT a_{\Box\mV\mV}$, $\deltaCT a_{H\mV\mV}$, $\deltaCT a_{d3}$ and $\deltaCT a_{H11}$; $\hat{\Gamma}_{WWHH}$ sets $\deltaCT b$, $\deltaCT a_{dd\mV\mV1}$, $\deltaCT a_{dd\mV\mV2}$, $\deltaCT a_{HHWW}$, $\deltaCT a_{HH11}$, $\deltaCT a_{H\Box{\cal V}{\cal V}}$, $\deltaCT a_{Hd2}$, $\deltaCT a_{Hd3}$ and $\deltaCT a_{HH\mV\mV}$; and with the singular parts of all the CTs, we check that $\hat{\Gamma}_{\pi WH}$ gives a finite contribution to the scattering amplitude.

Second, these divergent parts of the CTs can also be determined by using the renormalization conditions of \eqrefs{condOS1}{condOS5}. They allow us to write the counterterms as functions of the undressed 1PI functions. Then we have used this second procedure as a check of our results that we obtain solving the system described in the previous paragraph. 
Also,  with this second procedure we can determine the finite contributions to the counterterms (if any) and we use them in the final numerical computation of the  one-loop cross section in the next section.  Therefore,  we postpone the estimates of the finite contributions for the next section and focus here in the derivation of the singular parts of the EChL counterterms. 
For completeness, we also provide in \eqref{L2param-div} the divergent counterterms for the EW parameters derived in our previous work together with $\deltaCT\lambda$ (that enters now in the $s$ channel) in \eqref{L2param-div}.  
The corresponding SM results, obtained from the 1-leg and 2-legs Green functions, were presented and compared with the EChL in~\cite{Herrero:2021iqt} and we do not repeat them here. 

 
Our results for the divergent parts of the full set of EChL coefficients are then summarized as follows: 
\bear
&&\deltaCT a = \frac{\Delta_\epsilon}{16\pi^2}\frac{3}{2\vev^2}\left((a^2-b)(a-\kappa_3)\mh^2+a\left((1-3a^2+2b)\mw^2+(1-a^2)\mz^2\right)\right) \,,  \nn\\
&&\boldsymbol{\deltaCT b} = -\frac{\Delta_\epsilon}{16\pi^2}\frac{1}{2\vev^2}\left((a^2-b)(8a^2-2b-12a\kappa_3+3\kappa_4)\mh^2\right. \nn\\
&&\left.\hspace{30mm}+6a^2b(2\mw^2+\mz^2)-6b(\mw^2+\mz^2)-6b^2\mw^2\right) \nn\\
&&\boldsymbol{\deltaCT\kappa_3} = -\frac{\Delta_\epsilon}{16\pi^2}\frac{1}{2\mh^2\vev^2}\left(\kappa_3(a^2-b+9\kappa_3^2-6\kappa_4)\mh^4-3(1-a^2)\kappa_3\mh^2(\mw^2+\mz^2)\right.  \nn\\
&&\left. \hspace{15mm} +6(-2ab+2a^2\kappa_3+b\kappa_3)(2\mw^4+\mz^4)\right)\,, \nn\\
&&\boldsymbol{\deltaCT a_{dd\mV\mV 1}}=-\frac{\Delta_\epsilon}{16\pi^2}\frac{a^4+a^2b+b^2}{3},\qquad\boldsymbol{\deltaCT a_{dd\mV\mV 2}}=-\frac{\Delta_\epsilon}{16\pi^2}\frac{(a^2-b)(2a^2+b+6)}{12}, \nn\\
&&\deltaCT a_{11}=\frac{\Delta_\epsilon}{16\pi^2}\frac{a^2}{4} \,,\qquad\deltaCT a_{H11}=\frac{\Delta_\epsilon}{16\pi^2}\frac{a(a^2-b)}{2} \,,\qquad\boldsymbol{\deltaCT a_{HH11}}=\frac{\Delta_\epsilon}{16\pi^2}\frac{4a^4-5a^2b+b^2}{4}  \nn\\
&&\deltaCT a_{HWW} =\frac{\Delta_\epsilon}{16\pi^2}\frac{a(a^2-b)}{12} \,,\qquad\boldsymbol{\deltaCT a_{HHWW}}=-\frac{\Delta_\epsilon}{16\pi^2}\frac{4a^4-5a^2b+b^2}{24}\,,  \nn\\
&&\deltaCT a_{d2} = -\frac{\Delta_\epsilon}{16\pi^2}\frac{a(a^2-b)}{6}\,,\qquad\boldsymbol{\deltaCT a_{Hd2}}=\frac{\Delta_\epsilon}{16\pi^2}\frac{4a^4-5a^2b+b^2}{6}\,,  \nn\\
&&\deltaCT a_{\Box{\cal V}{\cal V}} = -\frac{\Delta_\epsilon}{16\pi^2}\frac{a(2+a^2)}{4} \,,  \qquad\boldsymbol{\deltaCT a_{H\Box{\cal V}{\cal V}}} = \frac{\Delta_\epsilon}{16\pi^2}\frac{4a^4+a^2(4-3b)-2b}{4}\,,  \nn\\
&&\deltaCT a_{d3} = \frac{\Delta_\epsilon}{16\pi^2}\frac{a(a^2+b)}{2}\,,\qquad\boldsymbol{\deltaCT a_{Hd3}}=\frac{\Delta_\epsilon}{16\pi^2}\frac{-4a^4+a^2b+b^2}{2}  \nn\\
&&\deltaCT a_{\Box\Box}=-\frac{\Delta_\epsilon}{16\pi^2}\frac{3a^2}{4} \,,\qquad\boldsymbol{\deltaCT a_{H\Box\Box}}=\frac{\Delta_\epsilon}{16\pi^2}\frac{3a(2a^2-b)}{2}\,,  \nn\\
&&\boldsymbol{\deltaCT a_{dd\Box}}=\frac{\Delta_\epsilon}{16\pi^2}\frac{3a(a^2-b)}{2}\,,\quad \boldsymbol{\deltaCT a_{Hdd}}=0\,,\quad \boldsymbol{\deltaCT a_{ddW}}/2=\boldsymbol{\deltaCT a_{ddZ}}=-\frac{\Delta_\epsilon}{16\pi^2}3a(a^2-b)\,,  \nn\\
&& \boldsymbol{\deltaCT a_{H\mV\mV}}=\boldsymbol{\deltaCT a_{HH\mV\mV}}=0\,, 
\label{L4param-div}
\eear
where we have used the $\textbf{bold}$ notation for the new EChL coefficients in this computation respect to~\cite{Herrero:2021iqt}.
As it is expected from the $SU(2)_L\times U(1)_Y$ gauge invariant construction of $\mL_4$, we have found  no $\xi$-dependence in any of the CTs of the EChL coefficients (in contrast to the results for the CTs of the EW parameters, like $\delta g$ etc that are in general $\xi$ dependent,  see~\cite{Herrero:2021iqt}). 
We also see in these results that 
some of these CTs vanish for the choice $a=b=\kappa_3=\kappa_4=1$, and some others do not,  like $a_{dd\mV\mV1}$, $a_{dd\mV\mV2}$, $a_{11}$, $a_{\Box{\cal V}{\cal V}}$, $a_{H\Box{\cal V}{\cal V}}$, $a_{d3}$, $a_{Hd3}$, $a_{\Box\Box}$ and $a_{H\Box\Box}$.  

Some comments about the previous results in \eqref{L4param-div} are in order.  First, we wish to notice that  these results, to our knowledge,  are the only ones within the EChL that apply to the most general and complete renormalization program of off-shell \1loop 1PI functions and including all types of bosonic loop diagrams in the $R_\xi$ gauges. 
However, it is pertinent to compare our results with some previous results of the EChL \1loop divergences and counterterms in the literature.  We will summarize this comparison in the following. 
Firstly,  we compare with previous works computing the one-loop scattering amplitude.  The renormalization of the $W_LW_L\to HH$ process  was studied to \1loop within the EChL previously in~\cite{ Delgado:2013hxa}.  It was done by means of the ET,  i.e., replacing the external $W_L$'s by the $w$ GB's and studying the corresponding  $ww \to HH$ scattering with just chiral loops (meaning loops with only GBs and Higgs in the internal propagators),  and assuming massless GBs (as in Landau gauge, i.e., for $\xi=0$). More recently in~\cite{Asiain:2021lch}, the loop contributions to the $W_L W_L \to HH$ scattering amplitude were computed as well by means of the ET,  i.e also for $ww \to HH$ scattering,  but improving  the previous computation of~\cite{ Delgado:2013hxa}, by considering all kind of bosonic \1loop diagrams  in this scattering of GBs. They also used the Landau gauge,  i.e. with massless GBs, and simplify the computation by assuming the so-called isospin limit with $\mw=\mz$.   We have further improved those two computations, in several aspects.    We do not use the ET,  i.e,  we consider gauge bosons in the external legs,  we work in generic $R_\xi$ gauges (i.e. with massive GBs) and we do not work in the isospin limit, i.e. for us $\mw$ and $\mz$ are different,  as corresponds to the physical on-shell gauge boson masses.   Furthermore,  we consider the full set of 1PI functions involved in the amplitude and include all kind of diagrams in those  functions.  The full set of one-loop diagrams  computed here are in consequence different than in~\cite{Asiain:2021lch}.  However, we can make contact with some of the results in this reference, by specifying our results for the particular assumptions and approximations of that reference.  For instance,  
taking into account the differences in the conventions,  and setting $\mz=\mw$,  we find agreement for the CTs of $a$, $b$, $\lambda$, $\kappa_3$, and $a_{d2}$.   On the other hand,  to compare with this reference,  it is convenient to use the reduced set of NLO coefficients that, as explained in the previous sections,  can be obtained by the use of the equations of motion. 
 Concretely,  the EChL NLO coefficients  appearing in the  scattering amplitude are those presented in 
 \eqref{ampWWtoHH-HEFT} and appear within the particular combinations of coefficients given in \eqref{delta-eta}.  Therefore these are the ones that should be compared with~\cite{Asiain:2021lch}.  
From our results in \eqref{L4param-div},  our prediction for the divergences of these combinations are:
\bear
\deltaCT\eta = \deltaCT \tilde{a}_{dd\mV\mV 1} = \deltaCT (a_{dd\mV\mV 1}-4a^2a_{11}+2aa_{d3}) &=&
- \frac{\Delta_\epsilon}{16\pi^2}\frac{(a^2-b)^2}{3} \,,  \nn\\ 
\deltaCT\delta = \deltaCT \tilde{a}_{dd\mV\mV 2} = \deltaCT (a_{dd\mV\mV 2}+\frac{a}{2}a_{dd\Box}) &=&
\frac{\Delta_\epsilon}{16\pi^2}\frac{(a^2-b)(7a^2-b-6)}{12}  \nn\\
\deltaCT (a_{H\mV\mV}-2a_{\Box\mV\mV}+2aa_{\Box\Box}) &=& \frac{\Delta_\epsilon}{16\pi^2}a(1-a^2)   \nn\\
\deltaCT (a_{HH\mV\mV}-6\kappa_3a_{\Box\mV\mV}-4a_{H\Box\mV\mV}+ \nn\\
+4ba_{\Box\Box}+6\kappa_3aa_{\Box\Box}+4aa_{H\Box\Box}) &=& \frac{\Delta_\epsilon}{16\pi^2}(3\kappa_3 a(1-a^2)+2b-2a^2(2+3b)+8a^4)    \nn\\
\deltaCT (a_{Hdd}-a_{dd\Box}) &=& -\frac{\Delta_\epsilon}{16\pi^2}\frac{3a(a^2-b)}{2}
\label{optildesCT}
\eear

The two first lines in the above equation are in agreement with the result for $\eta$ and $\delta$ in~\cite{Delgado:2013hxa,Asiain:2021lch},  where $a_{11}$, $a_{d3}$ and $a_{dd\Box}$ were not considered.  It is interesting to remark that the combinations in \eqref{optildesCT} indeed vanish for $a=b=\kappa_3=1$ as expected in the comparison with the SM. 

Secondly,  we compare our results with Ref.~\cite{Gavela:2014uta}.   In this work, the renormalization of \1loop 1PI functions was performed for off-shell external legs but  they considered the pure scalar theory, i.e., only the Higgs and GBs sector of the EChL and worked with massless GBs (as in Landau gauge, with $\xi=0$).  No gauge or ghost fields were included and, therefore,  no gauge-fixing.  We find agreement in the divergences found for the subset of $a_i$'s involved in the scalar sector (the coefficients in the notation of~\cite{Gavela:2014uta} are specified inside the parentheses). Concretely,  we agree in: $a$ ($a_C$), $b$ ($b_C$), $\kappa_3$ ($\mu_3$), $a_{dd\mV\mV1}$ ($c_8$), $a_{dd\mV\mV2}$ ($c_{20}$), $a_{11}$ ($c_9$), $a_{H11}$ ($a_9$), $a_{HH11}$ ($b_9$), $a_{dd\Box}$ ($c_{\Delta H}$), $a_{\Box{\cal V}{\cal V}}$ ($c_7$), $a_{H\Box{\cal V}{\cal V}}$ ($a_7$), $a_{d3}$ ($c_{10}$), $a_{Hd3}$ ($a_{10}$), $a_{\Box\Box}$ ($c_{\Box H}$) and $a_{H\Box\Box}$ ($a_{\Box H}$). 

Thirdly, we compare our results with others that do not study scattering amplitudes but are devoted to the renormalization of the Lagrangian.  In particular,  the renormalization of the EChL was studied in the path integral formalism,  using the background field method,  in~\cite{Guo:2015isa, Buchalla:2017jlu, Buchalla:2020kdh}.  The most complete comparison of our results should be done with the bosonic loop results of~\cite{Buchalla:2017jlu,Buchalla:2020kdh} since they also included all loops of scalar and gauge particles.  However, the comparison with the path integral results is tricky since they use the equations of motion to reduce the number of operators in the Lagrangian.  Therefore,  some off-shell divergences do not appear in their results and some others are redefined by the use of the equations of motion.  They also use redefinitions of the fields (in particular the Higgs field) to reach the canonical kinetic term in the Lagrangian.  On the other hand, the parametrization used in~\cite{Buchalla:2017jlu,Buchalla:2020kdh} is also very different than here and not straightforward to compare with. 
For example, the divergence canceled by ours $a_{ddW}$, $a_{ddZ}$ and $a_{Hdd}$ in the $HHH$ Green function is absorbed via the Higgs field redefinition in their context. 

 Finally,  we summarize in the following the main results regarding the renormalization group running equations (RGEs) for the  NLO EChL coefficients, which complement those given in our previous work \cite{Herrero:2021iqt}.  These RGEs can be easily derived from the previous results in \eqref{L4param-div} and taking into account the relation between the renormalized and  bare coefficients given  by $a^0_i=a_i+\delta a_i$.   In the $\overline{MS}$ scheme (with $\mu$ being the scale of dimensional regularization in $D=4 -\epsilon$ dimensions), the running  $a_i(\mu)$ can  be written as follows:
 \bear
a_i(\mu)& =& a_i^0 -\delta a_i(\mu) \,,\,\,\, \delta a_i(\mu)=\deltaCT a_i - \frac{\gamma_{a_i}}{16\pi^2}\log \mu^2 \,,\,\,\,
\deltaCT a_i =\frac{\Delta_\epsilon}{16\pi^2}\gamma_{a_i} \,, 
\label{airun}
\eear
where the divergent $\deltaCT a_i$ is written in terms of the anomalous dimension $\gamma_{a_i}$  of the corresponding effective operator. The running and renormalized $a_i$'s can then be related, in practice,  by:
 \bear
 a_i(\mu)& =&a_i+\frac{\gamma_{a_i}}{16\pi^2}\log \mu^2
 \label{airun-air}
 \eear
The set of RGEs for all the $a_i$'s then  immediately follow:
\bear
a_i(\mu)&=&a_i(\mu')+\frac{1}{16\pi^2}\gamma_{a_i}\log\left(\frac{\mu^2}{\mu'^2}\right) \,,
\label{RGE1}
\eear
where the specific value of $\gamma_{a_i}$ for each coefficient can be read from \eqref{L4param-div}.
For instance, in the case of the two most relevant NLO-EChL coefficients for the present $WW \to HH$ scattering,  $\eta$ and $\delta$,  we get the following RGE's:  
\bear
\eta(\mu)&=&\eta(\mu')-\frac{1}{16\pi^2}\frac{1}{3}(a^2-b)^2\log\left(\frac{\mu^2}{\mu'^2}\right) \,,  \nn\\
\delta(\mu)&=&\delta(\mu')+\frac{1}{16\pi^2}\frac{1}{12}(a^2-b)(7a^2-b-6)\log\left(\frac{\mu^2}{\mu'^2}\right) \,, 
\label{RGE-eta-delta}
\eear
which are in agreement with the RGEs given in~\cite{Delgado:2013hxa}.  
Notice that,  in particular,  for $a=b=1$ these two EChL coefficients $\eta$ and $\delta$ do not run,  therefore they are RGEs invariants.

\section{Numerical results for $W_L^+W_L^-\to HH$}
\label{sec-plots}
In this section we study the numerical predictions from the EChL for the cross section of the scattering process $WW \to HH$ and compare the tree level rates with the one-loop rates.  We also compare these rates with the SM case which have been computed independently here following the same procedure as for the EChL case.  It is also interesting the comparison of this SM case with previous SM results in the literature~\cite{Dreyer:2020xaj}.  Since, as we have already said,  the dominant contribution to this scattering process at the TeV domain is that coming from the longitudinally polarized  gauge boson modes, we will focus in this section in this most relevant cross section, i.e in  $\sigma(W_L^+W_L^-\to HH)$.  In addition,  this numerical study  of the radiative corrections will be devoted to  the most relevant coefficients of the NLO-EChL, that as already said are the parameters $\eta$ and $\delta$.   For simplicity,  the LO-EChL parameters will be set here to the SM default values, i.e.  in the following we set $a=b=\kappa_3=\kappa_4=1$.  All the numerical computations presented here have been performed with the help of FormCalc and LoopTools and, for definiteness,  we choose the Feynman 't Hooft gauge,  i.e we fix $\xi=1$. 

First of all,  it is worth mentioning that we have done a numerical check of the finiteness of the predicted one-loop cross section in both cases,  the EChL and the SM.  This is done indirectly,  by  checking numerically the renormalization $\mu$ scale independence of the result.  This is not a trivial  check at all,  since the computation of the one-loop amplitude from the 1PI functions amounts to the evaluation of more than 500 \1loop diagrams where each one depends on this $\mu$ scale.  Thus,  the cancellation of the $\mu$ dependence among the various diagrams found in the final result is a quite convincing check.  Notice that for the studied case here of $a=b=1$ the two parameters,  $\delta$ and $\eta$, as already said, do not run, therefore they have equal value at any assumed $\mu$ scale.  

\begin{figure}[t!]
\begin{center}
    \includegraphics[width=0.9\textwidth]{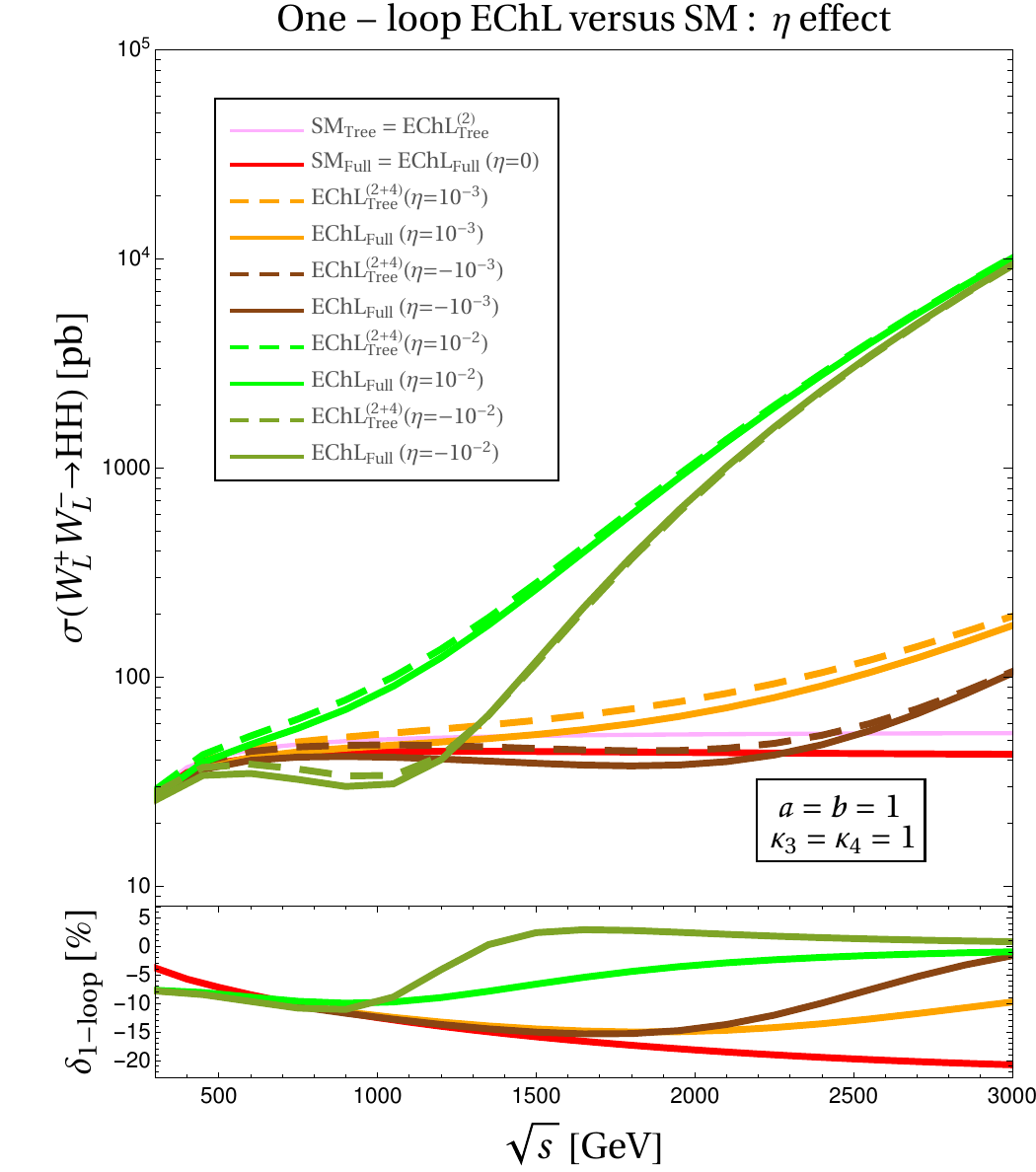}
\caption{Cross section prediction for $W_L^+ W_L^-\to HH$ as a function of the energy $\sqrt{s}$ within the EChL  at one-loop level (solid lines)  and comparison with the tree level prediction (dashed lines).  The effect of the NLO parameter $\eta$ is displayed,  assuming values for this parameter of $\pm 10^{-2}$ and $\pm 10^{-3}$.  The LO parameters are set to $a=b=\kappa_3=\kappa_4=1$.   The other NLO-parameters are set to zero.  The SM predictions at tree level (pink) and 1-loop level (red) are also included.  The relative size of the one-loop prediction respect to the tree level one,  defined by means of $\delta_{\rm 1-loop}$ in \eqref{delta1loop},  is displayed at the bottom of this figure.  The color code is: red (SM), orange (EChL, $\eta=10^{-3}$),  brown (EChL, $\eta=-10^{-3}$),  bright green (EChL, $\eta=10^{-2}$),  green (EChL, $\eta=-10^{-2}$). }
\label{plotEChL-etaonly}
\end{center}
\end{figure}

\begin{figure}[t!]
\begin{center}
    \includegraphics[width=0.9\textwidth]{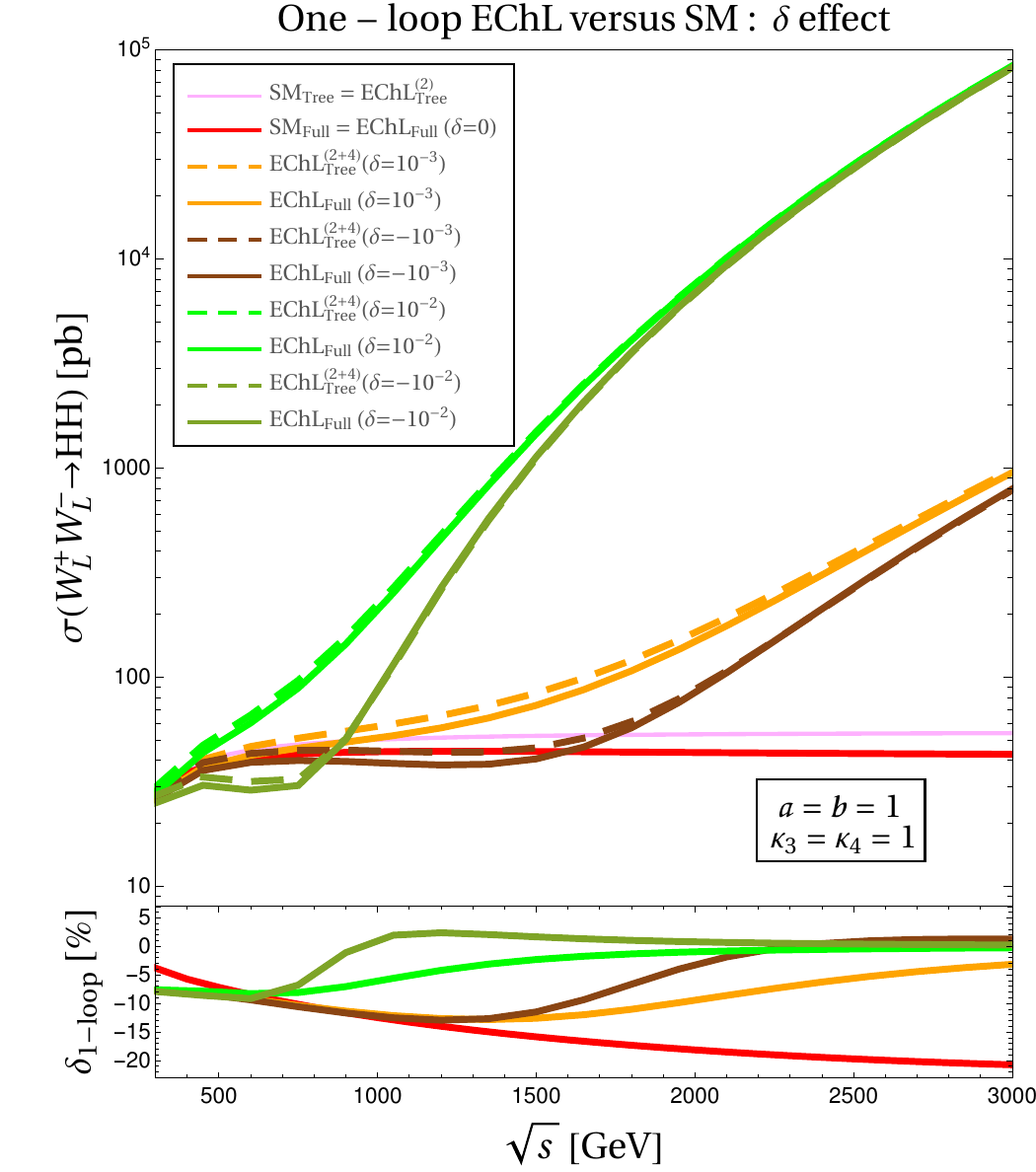}
\caption{Cross section prediction for $W_L^+ W_L^-\to HH$ as a function of the energy $\sqrt{s}$ within the EChL  at one-loop level (solid lines)  and comparison with the tree level prediction (dashed lines).  The effect of the NLO parameter $\delta$ is displayed,  assuming values for this parameter of $\pm 10^{-2}$ and $\pm 10^{-3}$.  The LO parameters are set to $a=b=\kappa_3=\kappa_4=1$.   The other NLO-parameters are set to zero.  The SM predictions at tree level (pink) and 1-loop level (red) are also included.  The relative size of the one-loop prediction respect to the tree level one,  defined by means of $\delta_{\rm 1-loop}$ in \eqref{delta1loop},  is displayed at the bottom of this figure.  The color code is: red (SM), orange (EChL, $\delta=10^{-3}$),  brown (EChL, $\delta=-10^{-3}$),  bright green (EChL, $\delta=10^{-2}$),  green (EChL, $\delta=-10^{-2}$). }
\label{plotEChL-deltaonly}
\end{center}
\end{figure}
We next summarize our numerical results for $\sigma(W_L^+W_L^-\to HH)$ as a function of the center-of-mass energy $\sqrt{s}$  in the two figures \ref{plotEChL-etaonly} and \ref{plotEChL-deltaonly}.  In \figref{plotEChL-etaonly} we study the effect of $\eta$,  and  in \figref{plotEChL-deltaonly} we study the effect of $\delta$.  In both cases, we have explored the following values for those coefficients,  $\pm 0.01$ and $\pm 0.001$,  which are  allowed by present experimental data.  In both plots we have included, for comparison,  the following rates: 1) the tree level predictions for the EChL,  ${\rm EChL}_{\rm Tree}^{(2+4)}$,  2) the full one-loop predictions for the EChL,  ${\rm EChL}_{\rm Full}$,  3)  the tree level predictions for the SM,  ${\rm SM}_{\rm Tree}$ (which coincide with the LO result in the EChL,  
${\rm EChL}_{\rm Tree}^{(2)}$),  and 4) the full one-loop predictions  for the SM.  In the lower part of these plots we display the predictions for the relative size of the one-loop correction respect to the tree level prediction, by means of $\delta_{\rm 1\, loop}$ that is defined by,
\be
\delta_{\rm{1-loop}}=\frac{(\sigma_{\rm Full}-\sigma_{\rm Tree})}{\sigma_{\rm Tree}}
\label{delta1loop}
\ee
The main features learnt from these two figures are the following:
\begin{itemize}
\item[$\star$] We get a one-loop correction in the SM case that is negative and increases in size with energy. The size of 
$\delta_{\rm{1-loop}}$ can be up to $\sim -20\%$ at the maximum energy studied of $\sqrt{s}=3$ TeV and it is in accordance with~\cite{Dreyer:2020xaj}.
\item[$\star$] The predictions from the EChL, both at tree level and one-loop level,  show a clear departure from the corresponding SM prediction.  The largest deviations  occur for the largest $|\delta|$ and/or $|\eta|$ considered values. 
\vspace{5cm}
\textcolor{white}{bla}

\textcolor{white}{bla}

\textcolor{white}{bla}

\item[$\star$] We get a one-loop correction in the EChL case that depending on the value of the coefficient and the value of the energy can be either negative or positive.  For $\eta$ we find it negative for $\pm 10^{-3}$ and $+10^{-2}$,  at all studied energies.  But it is positive for $-10^{-2}$ in the interval $1.3 {\rm TeV} < \sqrt{s} <3 {\rm TeV}$.  For $\delta$ we find it negative for $+10^{-2}$ and $+10^{-3}$ at all the studied energies.  But it is positive for $-10^{-2}$ in the interval $0.9 {\rm TeV} < \sqrt{s} <3 {\rm TeV}$ and for $-10^{-3}$ in the interval $2.5 {\rm TeV} < \sqrt{s} <3 {\rm TeV}$.
\item[$\star$]  Overall we see that the maximum size of the radiative one-loop correction found in the EChL is about $-15\%$ in both $\eta$ and $\delta$ cases.  This is a bit lower than in the SM case. 
\item[$\star$] Finally,  notice that the values of the coefficients  $\eta$ and  $\delta$ specified  in these plots refer to the renormalized parameter values.  However,  since we have taken in these plots,  $a=b=1$,  they do not depend on the $\mu$ scale.  This, together with the previously mentioned $\mu$ independence of the sum of all the contributing one-loop diagrams, complements the check of $\mu$ scale invariance of the total cross section result. 
\end{itemize}

\section{Conclusions}
\label{sec-conclu}
In this work we have computed the one-loop electroweak radiative corrections to the scattering process $WW \to HH$ within the context of the Higgs Effective Field Theory, considering that the new Higgs Physics beyond the SM enters only in the bosonic sector and it is given by the Electroweak Chiral Lagrangian.  We consider this EChL with all the relevant effective operators  of chiral dimensions two and four and present the computation in terms of the involved 1PI Green functions in covariant $R_\xi $ gauges.  An ambitious renormalization program for all these one-loop 1PI functions involved is developed, considering the most general case with arbitrary  momenta for the external particle legs.  This renormalization procedure  is more demanding than just requiring a finite result for the one-loop amplitude with external on-shell particles,  and it has the advantage of being applicable to several processes sharing some of those 1PI functions with the amplitude under study here.  We have applied this same procedure for both cases, the EChL and the SM.  In particular, we have used here for $WW \to HH$ scattering some of the previous renormalized 1PI functions computed in our previous work devoted to $WZ \to WZ$ scattering.  We have used those functions also here and then we have complemented them with the new one-loop 1PI functions for the new vertices involving the Higgs particle,  $HHH$, $HWW$,  $\pi WH$,  $AHH$, $ZHH$ and $WWHH$, whose results are presented here.    

One of the most important results contained in this work,  are the full set of divergent counterterms derived for the EChL coefficients, summarized in \eqrefs{L4param-div}{optildesCT}.  These set of divergences do also set the corresponding  set of  RGEs for the involved HEFT coefficients, according to \eqrefs{airun}{RGE-eta-delta}.  A small subset of  these results have been cross-checked with previous results in the literature which were done following a very different approach to ours and we have found agreement.  A discussion on this comparison has also  been included in the present work.

The final part of this paper has been devoted to the numerical computation of the one-loop radiative corrections to the cross section of the $W_LW_L \to HH$ scattering process.  Again we have done in parallel both the computation for the EChL and for the SM.  In the case of the SM we have found agreement with the previous result in~\cite{Dreyer:2020xaj}.  Our estimate of the one-loop correction respect the tree level cross section in the SM gives a negative value whose maximum size is reached at the largest energy studied of $\sqrt{s}=3$ TeV and is about $\delta_{\rm{1-loop}}\sim -20\%$.  In the EChL case,  where we have considered the effects from the two most relevant parameters $\eta$ and $\delta$,  we find also important one-loop corrections, with a maximum of about $\delta_{\rm{1-loop}}\sim -15\%$,  a bit lower than in the SM case.  The size of this correction depends on the energy and the particular values of the EChL coefficients.  The largest departures of the HEFT respect to the SM prediction are found for the largest studied values of $\delta$ and/or $\eta$.  There are also some input values for these parameters and energy ranges that provide a positive one-loop correction although small,  being below $5\%$.  All these numerical results are summarized in Figs. \ref{plotEChL-etaonly} and \ref{plotEChL-deltaonly}.

\section*{Acknowledgments}
We would like first to warmly thank Daniel Domenech for his valuable help in producing the figure 3.
The work of MH has received financial support from the ``Spanish Agencia 
Estatal de Investigaci\'on'' (AEI) and the EU
``Fondo Europeo de Desarrollo Regional'' (FEDER) through the project
   PID2019-108892RB-I00/AEI/10.13039/501100011033 and from the grant IFT
Centro de Excelencia Severo Ochoa SEV-2016-0597.  We also  acknowledge finantial support  from the European Union's Horizon 2020 research and innovation
programme under the Marie Sklodowska-Curie grant agreement No 674896 and
No 860881-HIDDeN,  he ITN ELUSIVES H2020-MSCA-ITN-2015//674896 and the RISE INVISIBLESPLUS H2020-MSCA-RISE-2015//690575.
The work of RM is supported by CONICET and ANPCyT under projects PICT 2016-0164, PICT 2017-2751 and PICT 2018-03682.
\newpage

\section*{Appendices}
\appendix
\section{Summary of complementary 1PI functions}
\label{App-previous-results}

For completeness, we summarize in this appendix the renormalized 1PI functions already derived in~\cite{Herrero:2021iqt} that also enter in the present $WW\to HH$ scattering.  We have taken these analytical results from that previous reference,  but  we have displayed them here by setting $\xi=1$,  as in the main results of this paper.  The definition of the EChL coefficients, parameters and functions entering in these complementary  functions can be found in \cite{Herrero:2021iqt}. 

Starting with the EChL, the 1-leg function (Higgs tadpole) is
\be
i\hat{T} = iT^{\rm Loop} -i\delta T\,,  \quad
\delta T = (\delta \mh^2 -\mh^2(-2\delta\Zf_H+\delta\lambda/\lambda+\Zf_\pi+2\delta\vev/\vev))\vev
\label{Htadpole}
\ee 
Notice that now we are fixing a typo in this counterterm respect to our previous publication.

The 2-legs functions are
\bear
-i\hat{\Sigma}_{HH}(q^2) &=& -i\Sigma_{HH}^{\rm Loop}(q^2) +i\left(\delta\Zf_H (q^2-\mh^2) -\delta\mh^2\right) +i\frac{2a_{\Box\Box}}{\vev^2}q^4 \,,  \nn\\
i\hat{\Sigma}_{WW}^T (q^2) &=& i\Sigma_{WW}^{T\,{\rm Loop}} (q^2) -i\left( \delta\Zf_W\left( q^2 -\mw^2 \right) -\delta\mw^2 \right)\,,  \nn\\
i\SER_{WW}^L (q^2) &=& i\Sigma_{WW}^{L\,{\rm Loop}} (q^2) +i\left( -\left( q^2 -\mw^2 \right)\delta\Zf_W +\delta\mw^2 +q^2\delta\xi_1\right) -iq^2g^2a_{11} \,, \nn\\
\SER_{W\pi} (q^2) &=& \Sigma_{W\pi}^{\rm Loop} (q^2) +\frac{\delta\xi_2-\delta\xi_1}{2}\mw^2 +q^2g^2a_{11} \,,  \nn\\
-i\SER_{\pi\pi} (q^2) &=& -i\Sigma_{\pi\pi}^{\rm Loop} (q^2) +i\left( \left( q^2 -\mw^2 \right)\delta\Zf_\pi -\delta\mw^2 -\mw^2\delta\xi_2 \right) -i\frac{g^2}{\mw^2}q^4a_{11}\,.
\label{SE-renorm}
\eear
In these formulas above, the  $a_i$ coefficients must be understood again as $a_i+\delta a_i$. 

On the other hand, 
the 3-legs functions corresponding to $W^\mu(k_1)W^\nu(k_2)V^\rho(q)$ (with $V=A,\,Z$) enter in the present work just at the LO, therefore,  they take the usual tree level expression:
\bear
i\Gamma_{W^+W^-A}^{\mu\nu\rho} &=& -ig\sw (g^{\mu\nu}(k_1-k_2)^\rho+g^{\nu\rho}(k_2-q)^\mu+g^{\rho\mu}(q-k_1)^\nu) \,,  \nn\\
i\Gamma_{W^+W^-Z}^{\mu\nu\rho} &=& -ig\cw (g^{\mu\nu}(k_1-k_2)^\rho+g^{\nu\rho}(k_2-q)^\mu+g^{\rho\mu}(q-k_1)^\nu) \,.
\label{3leg-renorm}
\eear
In contrast,  the $AHH$ and $ZHH$ 1PI functions in the second diagram of \figref{1PIdiagsWWHHinSTUC} vanish at LO and they get only NLO contributions that are finite.

Next we summarize the loops divergences of all the above 1PI functions.  These are: 
\bear 
iT^{\rm Loop}\vert_{div} &=& i\frac{\Delta_\epsilon}{16\pi^2} \frac{3} {2\vev}
\left(\kappa_3 \mh^4 +2a\left(2 \mw^4+\mz^4\right) \right)  \nn\\
\hspace{-3mm}-i\Sigma_{HH}^{\rm Loop}(q^2)\vert_{div} &=& i\frac{\Delta_\epsilon}{16\pi^2}\frac{3}{2\vev^2} 
\left(a^2q^4 -2a^2(2\mw^2+\mz^2)q^2 +(3\kappa_3^2+\kappa_4)\mh^4+(4a^2+2b)(2\mw^4+\mz^4)\right)  \nn\\
i\Sigma_{WW}^{T\,{\rm Loop}} (q^2)\vert_{div} &=& i\frac{\Delta_\epsilon}{16\pi^2}\frac{g^2}{12}\left((39-a^2)q^2 +3(a^2-b)\mh^2 +3(13-3a^2)\mw^2 -9\mz^2\right)  \nn\\
i\Sigma_{WW}^{L\,{\rm Loop}} (q^2)\vert_{div} &=& i\frac{\Delta_\epsilon}{16\pi^2}\frac{g^2}{4}\left(a^2q^2 +(a^2-b)\mh^2+(13-3a^2)\mw^2-3\mz^2\right) \nn\\
i\Sigma_{W\pi}^{\rm Loop} (q^2)\vert_{div} &=& i\frac{\Delta_\epsilon}{16\pi^2}\frac{g^2}{4}\left(-a^2q^2 -(a^2-b)\mh^2-(17/3-3a^2)\mw^2+(7/3)\mz^2\right)  \nn\\
-i\Sigma_{\pi\pi}^{\rm Loop} (q^2)\vert_{div} &=& i\frac{\Delta_\epsilon}{16\pi^2}\left(\frac{a^2}{\vev^2}q^4 +\frac{q^2}{\vev^2}((a^2-b)\mh^2-(5/3+3a^2)\mw^2-(5/3)\mz^2)\right) 
\label{old-div-Loop}
\eear
The resulting divergent part of the EChL counterterms for the EW parameters were also derived in~\cite{Herrero:2021iqt}.  We include those results here, including now explicitly $\deltaCT\lambda$, setting $\xi=1$:
\bear 
&&\deltaCT\Zf_H=\frac{\div}{16\pi^2}\frac{3a^2}{\vev^2}(2\mw^2+\mz^2)\,,\qquad \deltaCT T=\frac{\div}{16\pi^2} \frac{3}{2\vev}\left(\kappa_3\mh^4 +2a\left(2 \mw^4+\mz^4\right) \right)\,,  \nn\\
&&\deltaCT\mh^2=\frac{\div}{16\pi^2}\frac{3}{2\vev^2}((3\kappa_3^2+\kappa_4)\mh^4-2a^2\mh^2(2\mw^2+\mz^2)+(4a^2+2b)(2\mw^4+\mz^4))\,,  \nn\\
\vspace{1mm}
&&\deltaCT\Zf_B=-\frac{\div}{16\pi^2}\frac{\gY^2}{12}(1+a^2)\,, \qquad \deltaCT\Zf_W=\frac{\div}{16\pi^2}\frac{g^2}{12}(39-a^2)\,,  \nn\\
&&\deltaCT\mw^2=-\frac{\div}{16\pi^2}\frac{g^2}{12}\left(3(a^2-b)\mh^2 +(78-10a^2)\mw^2 -9\mz^2\right)\,,  \nn\\
    %
&&\deltaCT\mz^2=\frac{\div}{16\pi^2}\frac{g^2}{12\cw^2}\left(-3(a^2-b)\mh^2 +(7(1+a^2)+2(-43+a^2)\cw^2)\mw^2 +(10+a^2)\mz^2\right)\,,  \nn\\
&&\deltaCT\gY/\gY=0\,,  \qquad\deltaCT g/g=-\frac{\div}{16\pi^2}2g^2\,,  \nn\\
\vspace{1mm}
&&\deltaCT\xi_1=\frac{\div}{16\pi^2}\frac{g^2}{12}(39-a^2)\,,  \nn\\
&&\deltaCT\xi_2=\frac{\div}{16\pi^2}\frac{1}{3\vev^2}(6(a^2-b)\mh^2+(73-19a^2)\mw^2-14\mz^2)\,,  \nn\\
&&\deltaCT\Zf_\pi=-\frac{\div}{16\pi^2}\frac{1}{\vev^2}((a^2-b)\mh^2-(5/3+3a^2)\mw^2-(5/3)\mz^2)\,,  \nn\\
&&\deltaCT\vev/\vev=\frac{\div}{16\pi^2}\frac{2(\mw^2+\mz^2)}{3\vev^2}\,,  \nn\\
&&\deltaCT\lambda=\frac{\div}{16\pi^2}\frac{1}{4\vev^4} \left(2 a^2 \left(\mh^4+3 \mh^2 \left(\mw^2+\mz^2\right)+6 \left(2 \mw^4+\mz^4\right)\right)-6 a \left(2 \mw^4+\mz^4\right) \right.  \nn\\
&&\left.\hspace{13mm}-2 b \left(\mh^4-3 \left(2 \mw^4+\mz^4\right)\right)+3(3 \kappa_3^2 - \kappa_3+ \kappa_4) \mh^4-6 \mh^2 \left(\mw^2+\mz^2\right)\right) \,.
\label{L2param-div}
\eear

Finally the corresponding results in the SM with $\xi=1$ are:
\bear
i\hat{\overline{T}} &=& i\overline{T}^{\rm Loop} -i\delta \overline{T}\,,  \quad
\delta \overline{T} = (\delta \mh^2 -\mh^2(-\delta\Zf_\phi+\delta\lambda/\lambda+2\delta\vev/\vev))\vev \,,  \nn\\
-i\SERSM_{HH}(q^2) &=& -i\SMSigma_{HH}^{\rm Loop}(q^2) +i\left(\delta\Zf_\phi (q^2-\mh^2) -\delta\mh^2\right) \,,  \nn\\
i\SERSM_{WW}^T (q^2) &=& i\SMSigma_{WW}^{T\,{\rm Loop}} (q^2) -i\left( \delta\Zf_W\left( q^2 -\mw^2 \right) -\delta\mw^2 \right)\,,  \nn\\
i\SERSM_{WW}^L (q^2) &=& i\SMSigma_{WW}^{L\,{\rm Loop}} (q^2) +i\left( -\left( q^2 -\mw^2 \right)\delta\Zf_W +\delta\mw^2 +q^2\delta\xi_1\right) \,, \nn\\
\SERSM_{W\pi} (q^2) &=& \SMSigma_{W\pi}^{\rm Loop} (q^2) +\frac{\delta\xi_2-\delta\xi_1}{2}\mw^2 \,,  \nn\\
-i\SERSM_{\pi\pi} (q^2) &=& -i\SMSigma_{\pi\pi}^{\rm Loop} (q^2) +i\left( \left( q^2 -\mw^2 \right)\delta\Zf_\phi -\delta\mw^2 -\mw^2\delta\xi_2 -\delta \overline{T}/\vev\right) \,,  
\eear
and again the $WWA$ and the $WWZ$ vertices enter only at the tree level in this amplitude, therefore:
\bear
i\overline{\Gamma}_{W^+W^-A}^{\mu\nu\rho} &=& -ig\sw (g^{\mu\nu}(k_1-k_2)^\rho+g^{\nu\rho}(k_2-q)^\mu+g^{\rho\mu}(q-k_1)^\nu) \,,  \nn\\
i\overline{\Gamma}_{W^+W^-Z}^{\mu\nu\rho} &=& -ig\cw (g^{\mu\nu}(k_1-k_2)^\rho+g^{\nu\rho}(k_2-q)^\mu+g^{\rho\mu}(q-k_1)^\nu) \,, 
\eear
whereas the $AHH$ and  $ZHH$ vertices vanish at the tree level and these 1PI functions only get 1-loop  corrections that are finite. 

The loop divergences of the above 1PI functions in the SM are:
\bear
i\overline{T}^{\rm Loop}\vert_{div} &=& i\frac{\Delta_\epsilon}{16\pi^2}\frac{1}{2\vev} 
 (3\mh^4 +6\left(2 \mw^4+\mz^4\right)+\mh^2(2 \mw^2+\mz^2)) \nn\\
-i\SMSigma_{HH}^{\rm Loop}(q^2)\vert_{div} &=& i\frac{\Delta_\epsilon}{16\pi^2}
\frac{1}{2\vev^2} 
( -4(2\mw^2+\mz^2)q^2 +15\mh^4+18(2\mw^4+\mz^4)+\mh^2(2 \mw^2+\mz^2))  \nn\\
i\SMSigma_{WW}^{T\,{\rm Loop}} (q^2)\vert_{div} &=& i\frac{\Delta_\epsilon}{16\pi^2}\frac{g^2}{6}\left(19q^2 +6(2\mw^2 -\mz^2)\right)  \nn\\
i\SMSigma_{WW}^{L\,{\rm Loop}} (q^2)\vert_{div} &=& i\frac{\Delta_\epsilon}{16\pi^2}g^2(2\mw^2-\mz^2) \nn\\
i\SMSigma_{W\pi}^{\rm Loop} (q^2)\vert_{div} &=& i\frac{\Delta_\epsilon}{16\pi^2}\frac{g^2}{4}\left(-2\mw^2+3\mz^2\right)  \nn\\
-i\SMSigma_{\pi\pi}^{\rm Loop} (q^2)\vert_{div} &=& i\frac{\Delta_\epsilon}{16\pi^2}\frac{g^2}{2}\left(-4(2\mw^2+\mz^2)q^2+3\mh^2+12\mw^4+6\mz^4+\mh^2(2\mw^2+\mz^2)\right)
\eear
and the resulting divergences of the counterterms are:
\bear
&&\deltaCT\Zf_\phi=\frac{\div}{16\pi^2}\frac{2}{\vev^2}(2\mw^2+\mz^2)\,,\quad \deltaCT \overline{T}=\frac{\div}{16\pi^2} \frac{1}{2\vev}\left(3\mh^4 +6\left(2 \mw^4+\mz^4\right)+\mh^2(2\mw^2+\mz^2) \right)\,,  \nn\\
&&\deltaCT\mh^2=\frac{\div}{16\pi^2}\frac{3}{2\vev^2}(5\mh^4-\mh^2(2\mw^2+\mz^2)+6(2\mw^4+\mz^4))\,,  \nn\\
&&\deltaCT\Zf_B=-\frac{\div}{16\pi^2}\frac{\gY^2}{6}\,, \qquad \deltaCT\Zf_W=\frac{\div}{16\pi^2}\frac{19g^2}{6}\,,  \nn\\
&&\deltaCT\mw^2=-\frac{\div}{16\pi^2}\frac{g^2}{6}\left(31\mw^2 -6\mz^2\right)\,,  \nn\\
&&\deltaCT\mz^2=\frac{\div}{16\pi^2}\frac{g^2}{6\cw^2}\left((10-42\cw^2)\mw^2 +7\mz^2\right)\,,  \nn\\
&&\deltaCT\gY/\gY=0\,,  \qquad\deltaCT g/g=-\frac{\div}{16\pi^2}2g^2\,,  \nn\\
&&\deltaCT\xi_1=\frac{\div}{16\pi^2}\frac{19g^2}{6}\,,  \qquad\deltaCT\xi_2=\frac{\div}{16\pi^2}\frac{2}{3\vev^2}(25\mw^2-9\mz^2)\,,  \nn\\
&&\deltaCT\vev/\vev=\frac{\div}{16\pi^2}\frac{2\mw^2+\mz^2}{\vev^2}\,,  \nn\\
&&\deltaCT\lambda=\frac{\div}{16\pi^2}\frac{1}{\vev^4}(3 \mh^4-\mh^2 \left(2 \mw^2+\mz^2\right)+3 \left(2 \mw^4+\mz^4\right)) \,.
\label{SMparam-div}
\eear

\section{Relevant \1loop diagrams}
\label{App-oneloopdiag}
In this Appendix we present the relevant \1loop diagrams entering in the computation of the 1PI functions for $WW \to HH$ scattering  within the EChL.   
In particular, the corresponding ones to the new Green functions, $\Gamma_{HHH}$, $\Gamma_{\pi WH}$, $\Gamma_{AHH}$, $\Gamma_{ZHH}$ and $\Gamma_{WWHH}$, which respect to our previous computation in~\cite{Herrero:2021iqt}.
These diagrams were generated with FeynArts~\cite{FeynArts} and we collect them by different topologies using a generic notation for the internal propagators: dashed lines refer to both  Higgs  boson or Goldstone bosons and  wavy lines refer to all possible EW gauge bosons. Notice the absence of ghost fields since the Higgs boson does not interact with them in the EChL, but they are present in the SM computation. 

\begin{figure}[H]
\begin{center}
    \includegraphics[width=0.95\textwidth]{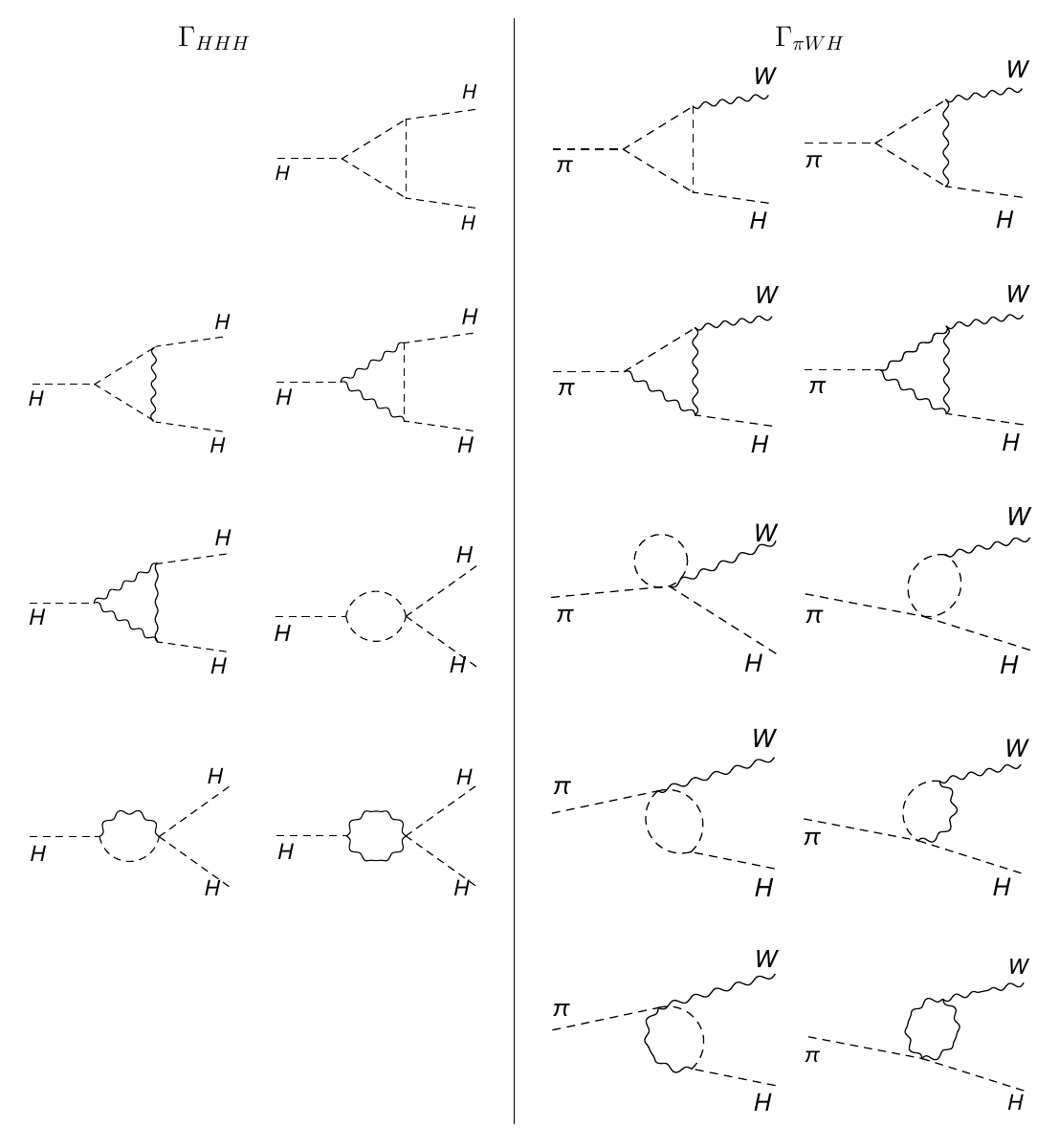}
\caption{Generic loop diagrams for the $HHH$ and $\pi WH$ Green functions in the EChL. The topologies for $AHH$ and $ZHH$ are the same as for $\pi WH$.}
\label{new3legs-generic-loops}
\end{center}
\end{figure}

The loop diagrams of $\Gamma_{HHH}$ are shown in the first column of \figref{new3legs-generic-loops}.  Different from the SM, the results in the EChL depend on $a$,  $b$,  $\kappa_3$ and $\kappa_4$  and there is a different (non trivial) momentum dependence due to the behaviour of the scalar loop diagrams in the EChL and the SM.
The same conclusions for the diagrams in the second column corresponding to $\Gamma_{\pi WH}$, but there is no $\kappa_4$ dependence here.

Regarding the $AHH$ and $ZHH$ Green functions, they have the same generic topologies than $\Gamma_{\pi WH}$ but they result finite in both the EChL and SM.  We omit the corresponding diagrams for shortness.

\begin{figure}[H]
\begin{center}
    \includegraphics[width=.95\textwidth]{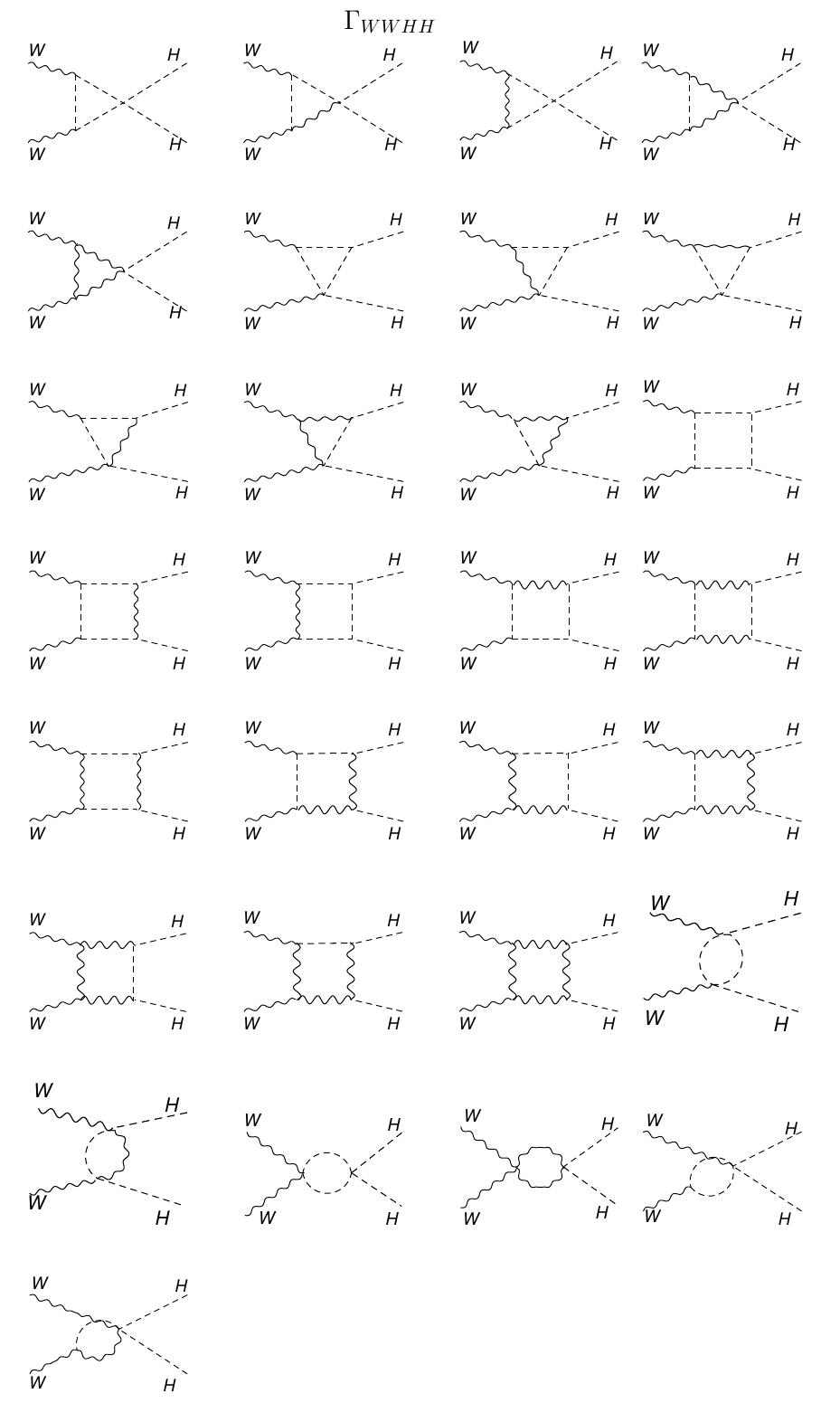}
\caption{Generic loop diagrams for the $WWHH$  Green functions in the EChL.}
\label{WWHH-generic-loops}
\end{center}
\end{figure}

Finally the \1loop diagrams for the $WWHH$ 1PI Green function are presented in \figref{WWHH-generic-loops}.  Also, the results in the EChL depend on $a$, $b$,  $\kappa_3$ and $\kappa_4$ and again there is a
different (non trivial) momentum dependence.

\bibliography{HM2-arXiv-v2}
\end{document}